\documentclass[aps,prb]{revtex4}
\usepackage{amsfonts}
\usepackage{graphicx}
\usepackage{psfrag}
\usepackage{epsf}
\usepackage{epsfig}
\def\bea{\begin{eqnarray}}
\def\eea{\end{eqnarray}}
\def\be{\begin{equation}}
\def\ee{\end{equation}}

\begin{document}
\title{The Kondo screening cloud: what it is and how to observe it}
\author{Ian Affleck}
\affiliation{Department of Physics and Astronomy, University of British
  Columbia, Vancouver, BC, Canada V6T 1Z1}
\begin{abstract}The Kondo effect involves the formation of a spin singlet 
by a magnetic impurity and conduction electrons.  It is  characterized 
by a low temperature scale, the Kondo temperature, $T_K$, and an 
associated long length scale, $\xi_K\equiv 
\hbar v_F/(k_BT_K)$ where $v_F$ is the 
Fermi velocity. This Kondo length is often estimated theoretically 
to be in the range of .1 to 1 microns but such a long characteristic 
length scale has never been observed experimentally. In this review, 
I will examine how $\xi_K$ appears as a crossover scale when one 
probes either the dependence of physical quantities an distance from 
the impurity or when the impurity is embedded in a finite size structure 
and discuss possible experiments that might finally observe this 
elusive length scale.
\end{abstract}
\maketitle

\section{Introduction\label{sec:intro}}
The Kondo model gives a simplified description of the interaction of 
a single magnetic impurity with the conduction electrons in a host, with 
Hamiltonian:
\begin{equation}
H-\mu N=\sum_{\vec k\sigma}\Psi^\dagger_{\vec k\sigma}\Psi_{\vec k\sigma}
\epsilon_k + J\vec S_{\hbox{imp}}\cdot \vec S_{\hbox{el}}(r=0).
\label{HK}\end{equation}
Here $\vec S_{\hbox{imp}}$ is the impurity spin operator 
and $\vec S_{\hbox{el}}$ is the electron spin density at position $\vec r$:
\be \vec S_{\hbox{el}}(\vec r)\equiv \Psi^\dagger_\alpha  {\vec 
\sigma_{\alpha \beta} \over 2}
\Psi_\beta (\vec r)\ee
where $\Psi_\alpha (\vec r)$ annihilates a conduction electron of spin $\alpha$ 
at $\vec r$ and repeated spin indices are summed over. We will focus 
on the case $S_{\hbox{imp}}=1/2$ although much of the discussion here carries 
over to higher spin. The dispersion relation will often be assumed to
be that of free electrons:
\be \epsilon_k={k^2\over 2m}-\epsilon_F,\ee
although that is not crucial.  (For a general 
review of the Kondo model, see Ref.~[\onlinecite{Hewson}].) 
Essentially the same model can be used to describe a quantum dot 
in the Coulomb blockade regime in a spin 1/2 state, interacting 
with one or more metallic leads (which can be formed by a 2 dimensional 
electron gas or by quantum wires in a semi-conductor inversion layer).

The dimensionless parameter which measures the strength of the Kondo 
interaction is $J\nu\equiv \lambda_0$ where $\nu$ is the density of states at the 
Fermi energy. For the free electron dispersion in D dimensions, this is:
\be \nu_D=k_F^{D-1}/(c_Dv_F)\ee
where $c_3=2\pi^2$, $c_2=2\pi$ and $c_1=\pi$. 
 Typically $J\nu \ll 1$ suggesting that perturbation theory 
could be useful. However, perturbation theory encouters infrared divergences 
at low energy scales.  In general, an $n^{th}$ order term in perturbation 
theory for some physical quantity characterized by an energy $E$ is 
proportional to $[J\nu \ln ({\cal D}/E)]^n$, where ${\cal D}$ is an ultra-violet 
scale of order the bandwidth or Fermi energy. It is found that the leading 
logarithmic divergences can be summed by expressing perturbation theory
in terms of the renormalized coupling constant at scale $E$:
\be \lambda (E)\approx \lambda_0 +\lambda_0^2\ln ({\cal D}/E)+\ldots \ee
Note that $\lambda (E)$ increases in magnitude, as $E$ is lowered, 
assuming it is initially positive (antiferromagnetic). $\lambda (E)$ 
obeys the renormalization group equation:
\be{d\lambda \over d\ln E}=\beta (\lambda )
=-\lambda (E)^2+(1/2)\lambda (E)^3+\ldots 
\label{RGE}\ee
where $\beta (\lambda )$ is the $\beta$-function. 
Keeping only the quadratic term in this RG equation, we find:
\be \lambda (E)\approx {\lambda_0 \over 1-\lambda_0 \ln ({\cal D}/E)}\label{lambdaE}\ee
Note that $\lambda_0$, the bare coupling, is the value of the 
renormalized coupling at the ultraviolet
scale, ${\cal D}$. As the energy scale is lowered, $\lambda$  
becomes of O(1) at the Kondo energy scale:
\be k_BT_K={\cal D}e^{-1/\lambda_0 }.\ee 
This renormalization group equation, (\ref{RGE}) is derived 
by integrating out high energy Fourier modes, reducing the effective 
bandwidth from $O(D)$, in energy units, to $O(E)$. Once the 
bandwidth becomes narrow (as happens at low energies) the width 
in wave-vector is related to the width in energy by $E=v_F\hbar |k-k_F|$ 
where $v_F$ is the Fermi velocity. Thus we may equally well describe 
$\lambda (E)$ as the effective coupling at wave-vector scale 
$E/(\hbar v_F)$ as at energy $E$. From this perspective,
 $k_BT_K/(\hbar v_F)$ is 
the characteristic wave-vector scale at which the effective Kondo 
coupling becomes large. Thus it is natural to introduce the Kondo 
length scale:
\be \xi_K\equiv \hbar v_F/(k_BT_K)\ee
(We henceforth set $k_B$ and $\hbar$ to one.) It seems appropriate 
to think of the effective Kondo coupling as growing at large distances, 
becoming large at length scales of order $\xi_K$. Equivalently, 
we might expect physical quantities depending on a length scale $r$ 
 to be scaling functions of the ratio $r/\xi_K$ rather than depending 
on $r$ and $\xi_K$ separately. 

There is an interesting analogy here with Quantum Chromodynamics (QCD)
the theory of the strong interactions in high energy physics. The 
renormalization group equation for the effective QCD coupling, $g_{QCD}$, 
(describing the interactions of gluons with themselves and with quarks) 
is the same as Eq. (\ref{RGE}) at quadratic order. It also gets 
large as the energy scale is lowered, and small as it is raised. 
In this case, one switches back and forth from energy units 
to momentum units using the velocity of light, $c$. The characteristic 
energy scale, $\Lambda_{QCD}$ where $g_{QCD}$ becomes $O(1)$ is of 
order 1 GeV, the mass of the proton, and the corresponding length 
scale is or order the Compton wavelength of the proton, 1 Fermi 
or $10^{-15}$ m. If the quarks inside a proton are probed with 
high energy (short wavelength) 
photons they appear nearly free.  On the other hand, 
they exhibit confinement (with an interaction growing linearly 
with separation) at long distances. In high energy physics 
it is commonplace to go back and forth freely between energy 
and length units using $c$ (and $\hbar$). However, there are 
well understood pitfalls in taking this picture of 
the effective coupling getting weak at high energies and strong 
at low energies, too literally. (These pitfalls occur both 
in the energy and distance picture, and do not seem to be particularly 
related to the distance viewpoint.) In this review article, 
we will be concerned with the validity of the corresponding picture 
for the Kondo model. 

The $\delta$-function form of the Kondo interaction implies that 
only the s-wave harmonic interacts with the impurity so that the 
model becomes fundamentally 1-dimensional. After linearizing the 
dispersion relation about the Fermi surface, the low energy 
effective Hamiltonian becomes:
\begin{equation}
H={iv_F\over 2\pi}\int_0^\infty dr\left[ \psi^\dagger_L{d\over dr}\psi_L
-\psi^\dagger_R{d\over dr}\psi_R\right] + v_F\lambda 
\vec S_{\hbox{imp}}\cdot \vec S_{\hbox{el}}(0).\label{H1D}\end{equation}
Here $r$ is the radial coordinate and $\psi_{L/R}$ represent 
incoming and outgoing waves, with the boundary condition:
\be \psi_L(0)=\psi_R(0).\label{freebc}\ee
These are defined in terms of the s-wave part of the 3D electron annihilation operator, $\Psi (\vec r)$ by:
\be \Psi (\vec r)={1\over \sqrt{2}\pi r}
\left[e^{-ik_Fr}\psi_L(r)-e^{ik_Fr}\psi_R(r)\right]+\ldots \label{3D1D}\ee
where the $\ldots$ represents higher spherical harmonics. 
Note that we have normalized the fermion fields as in  Ref.~[\onlinecite{AL}] 
so that:
\be \{\psi_L^\dagger (r),\psi_L(r')\}=2\pi \delta (r-r').\ee
In the limit of zero Kondo interactions, $\lambda_0 =0$, 
this describes a relativistic Dirac fermion with the Fermi velocity
playing the role of the velocity of light.  In such a model 
it is natural to go back and forth between energy and length 
units using $v_F$. Note that it is crucial to this estimate 
of $\xi_K$  that only 
one channel (the s-wave) couples to the Kondo impurity, allowing a 
mapping into a 1D model. While this 
happens in a variety of circumstances, including the case of a
quantum dot coupled to 2DEG's, there are also important cases where it 
can fail, which will be discussed later. 

An intuitive picture of this Kondo length scale can be obtained 
from considering the low energy strong coupling behavior of the model. 
This is most easily understood from a
1 dimensional tight-binding version of the model with Hamiltonian:
\begin{equation}
H=-t\sum_{j=0}^\infty (c^\dagger_jc_{j+1}+h.c.)+JS_{\hbox{imp}}
\cdot \vec S_{\hbox{el}}(0).\end{equation}
The strong coupling limit is easily understood.  When $J\gg t>0$, 
one electron gets trapped at the origin to form a singlet with the 
impurity spin. Note that this ``Kondo screening'' actually corresponds to the 
formation of an entangled state between the impurity spin and one 
conduction electron. The other electrons can do whatever they want 
except that they cannot enter or leave the origin since this would 
break up the Kondo singlet, costing an energy of $O(J)$. They 
effectively feel an infinite repulsion from the origin, corresponding 
to a $\pi /2$ phase shift. This is simply a boundary condition on 
otherwise free electrons. 

While the strong coupling limit is trivial, 
we are actually interested in the case of weak bare coupling, at low 
energies and long distances, where the effective coupling becomes 
strong. It is known from various approximate and exact calculations 
that the low energy physics (at $E\ll T_K$) is described by a local spin singlet
state and non-interacting 
electrons with a $\pi /2$ phase shift. On the other hand, the 
physics at intermediate energy scales, of $O(T_K)$ is complicated. 
To form a spin singlet with an S=1/2 impurity, only one electron is needed 
and an intuitive picture is that one electron is ``removed from the Fermi 
sea'' for this purpose.  However, unlike the simple case of large 
bare coupling, it would be quite wrong to think that this electron 
is localized at the origin.  The natural length scale over which we 
may think of this electron's wave-function being non-negligible is $\xi_K$. 
Such a naive picture must be used with caution. At best, 
it is valid only at long distances and low energies. If we probe 
the screened impurity with a long wavelength probe, this picture may 
apply. At shorter distances, it certainly breaks down. 
While this is only an intuitive picture, it nonetheless seems reasonable 
that $\xi_K$ will appear as a characteristic  scale  in any distance  
dependent physical property of a Kondo system.  The nature of the 
crossover at $\xi_K$ is the subject of this review.  

Nozi\`eres' local Fermi liquid theory is well-known to provide
 a powerful way of 
studying the behaviour of distance-independent quantities at 
low $T\ll T_K$. It turns out to also be useful for studying 
distance-dependent quantities at large distances $r\gg \xi_K$. 
In this approach, the screened impurity spin is eliminated from 
the effective Hamiltonian which, in lowest approximation, just 
consists of non-interacting electrons with a $\pi /2$ phase shift. 
Then, interactions at the origin are added for these phase shifted 
electrons, which originate from virtual excitations of the 
screened impurity complex. We may simply list all possible 
interactions at the origin allowed by symmetry. Assuming 
exact particle-hole symmetry, there is only one important 
leading irrelevant interaction:
\be H_{\hbox{int}}=-{v_F^2\over 6T_K}\vec J_L^2(0).\label{HFL}\ee
Here the current operator is defined as:
\be \vec J_L(r)\equiv  \psi_L^\dagger (r){\vec \sigma \over 2}\psi_L(r).\ee
$H_{\hbox{int}}$ contains a dimension 2 operator, so its coupling constant must 
have  dimensions of inverse energy. On general grounds, we 
expect that this coupling constant will be of order the crossover 
scale, $1/T_K$. The factor of $1/6$ in Eq. (\ref{HFL}) is 
just a matter of convenience. With this normalization, 
a calculation of the impurity susceptibility, to first order in 
this interaction gives:
\be \chi (T)\to {1\over 4T_K}.\ee
[This factor of 1/4 in $\chi (T)$ is conventional since in the 
high $T\gg T_K$ limit we obtain the free impurity result, 
$\chi (T)\to 1/(4T)$.] We emphasize that various other 
definitions of $T_K$ are in common use, differing by factors of O(1). 

Perhaps surprisingly, distance dependent aspects of Kondo physics 
are much less well-studied than energy dependent ones. At a theoretical 
level this may be a consequence of the fact that it is very difficult 
to get distance dependence from two of the most powerful methods 
of studying the Kondo effect: the Numerical Renormalization Group (NRG)
method, introduced by Wilson,\cite{Wilson} and the Bethe ansatz (BA) solution, 
discovered by Andrei\cite{Andrei} and Weigmann\cite{Weigmann}.
 The difficulty with NRG lies 
in the ``logarithmic discretization'' or the fact that an effective 
1D tight binding model is introduced in which the hopping parameter 
decays exponentially with distance from the impurity. This is, in fact, 
only an approximation to the full problem and one which seems to 
fail to keep track of length-dependence properly. However, more recently 
a way around this difficulty is being exploited\cite{Borda,ABS} which involves 
a more complicated and numerically costly version of NRG. The difficulty 
with BA seems to be that while exact wave-functions are calculated, 
they are given in such a complicated form that it is very challenging 
to calculate any Green's functions with them. Thus much of our 
understanding of length dependence comes from perturbation theory 
and analytic RG arguments, various mean field theories, exact 
diagonalization of short systems 
and Density Matrix Renormalization Group (DMRG) numerical results 
which are also restricted to fairly short systems (up to 32 sites).
The reasons why this exponentially large scale, $\xi_K=v_F/T_K$, has 
never been seen experimentally are probably that the associated 
crossover effects at this length scale 
are rather weak and subtle and that the basic Kondo model may 
not be adequate to describe the experimental systems used. 
Some inprovements to the model that might be neccessary are: including 
charge fluctuations (as in the Anderson model) taking into account 
a finite density of magnetic impurities, taking into account 
a finite density of non-magnetic impurities and taking into account 
electron-electron interactions even away from the impurity location. 
Thus  searches for effects at this length scale 
represent both an experimental challenge to find sufficiently 
ideal systems and also an opportunity to study the limits of 
validity of the basic model.

We will be concerned with two types of length dependence. The first 
type, analysed  in Sec. \ref{sec:inf}, involves a single impurity in an infinite host,
 described by 
the Hamiltonian of Eq. (\ref{HK}). We consider various 
observables as a function of distance from the impurity. 
The first one (Sub-Section \ref{sec:Knight}) is the Knight shift, measurable in nuclear magnetic 
resonance (NMR) experiments. This is simply the magnetic 
polarization of the electrons as a function of distance from the 
impurity, in linear response to a magnetic field (applied to 
both the impurity and the conduction electrons). The second (Sub-Section \ref{sec:Friedel}) 
is the charge density (Friedel) oscillations, as a function of 
distance from the impurity.  These could be observed by 
scanning tunneling microscopy (STM) for a magnetic impurity on a 
metallic surface.  The third (Subsection \ref{sec:corr})
 is the ground state equal time correlation function of the impurity spin and
 electron spin 
density at distance $r$ from the impurity. While probably not 
experimentally observable, this has a rather direct interpretation 
in terms of measuring the spatial probability distribution for the electron 
forming the spin singlet with the impurity. The second 
type of length dependence, analysed in Sec. \ref{sec:meso},
 occurs in mesocopic samples containing 
a single localized S=1/2 where the size of some part of the device 
is comparable to $\xi_K$. Here we consider four different 
situations. Two of them involve a S=1/2 quantum dot coupled to a 
finite length one-dimensional 
quantum wire.  In the first case, (Subsection \ref{sec:pers}) this wire is closed into a ring 
and we consider the persistent current through it as a function 
of the ring length and magnetic flux. In the second case, (Subsection \ref{sec:steps}) the ring 
is straight with the quantum dot coupled to one end and the 
other end open. In Subsection \ref{sec:transport}  we discuss  an S=1/2 quantum dot embedded in the centre of a finite length 
quantum wire which is tunnel-coupled to leads. 
In Subsection \ref{sec:box}, we contrast these strictly 1D models with the case of a magnetic 
impurity inside a three dimensional magnetic sample 
containing non-magnetic disorder: the Kondo box model.\cite{Thimm} 
Given certain simplifying assumptions about the disorder, the 
characteristic Kondo length scale is much smaller than $v_F/T_K$. 
It turns out that this is not just a consequence of the 3 dimensionality 
but also of the disorder.   In an ideal 3D Kondo box $\xi_K$ would 
again be the characteristic length, as we discuss.   Electron-electron 
interactions away from the impurity are ignored in all sections except \ref{sec:steps}, 
a quantum dot at the end of a finite quantum wire, where they are included 
using Luttinger liquid techniques. Sec. \ref{sec:conclude} 
contains conclusions.  

\section{Impurity in an infinite host\label{sec:inf}}
In this section, we consider the Hamiltonian of Eq. (\ref{HK}) for a 
single impurity interacting with electrons in an infinite volume. This 
is the traditional model for magnetic impurities in a metal. Note, however, 
that we ignore the presence of other impurities. A simplifying feature 
of this situation is that the mapping onto a single channel 1D system is exact.
 As we will see in sub-section \ref{sec:box}, 
this is not generally the case when we put a 3D system 
into a finite box.  Reflections off the boundary of the ``box'' greatly complicate
the model. 

An immediate question that arises is: how dilute must the magnetic 
impurities be to justify a single impurity approximation?  Given 
the screening cloud picture, one might come to the pessimistic 
conclusion that the average separation of impurities should be 
much larger than $\xi_K$:
\be n_{\hbox{imp}}^{-1/3}\gg \xi_K ??\ee
(Here $n_{\hbox{imp}}$ is the impurity density.) 
Such low densities are rarely, perhaps never, acheived in experiments 
on metals with dilute concentrations of magnetic impurities. Since 
typical ratios of $\xi_K$ to lattice constant can be in the thousands 
this would require densities $n_{\hbox{imp}}\ll 10^{-9}$ per unit cell.  
The reason\cite{BA3} that the condition is much weaker than this 
is basically that screening cloud wave-functions around different 
impurities become nearly orthogonal even when their centres are 
much closer together than their size. To see this consider 
two spherically symmetric wave-functions, centred around the points 
$\pm \vec R/2$ but otherwise identical,
 only containing Fourier modes within a narrow 
band of wave-vectors of width $\xi_K^{-1}$ around $k_F$. The overlap is:
\be O(R)=\int d^3r\psi^*(|\vec r+\vec R/2|)\psi (|\vec r-\vec R/2|).\ee
In terms of the Fourier transform of the screening cloud wave-function, 
this becomes:
\be O(R)=\int {d^3k\over (2\pi )^3}e^{i\vec k\cdot \vec R}|\psi (k)|^2
=\int_0^\infty dkk^2|\psi (k)|^2{\sin (kR)\over 2\pi^2kR}.\ee
We expect the screening cloud wave-function to decay on the length 
scale $\xi_K$ but to oscillate at the Fermi wave-vector. It should 
be built out of wave-vectors within $\xi_K^{-1}$ of $k_F$.  Thus we may
assume:
\be |\psi (k)|^2\approx (\xi_K/k_F^2)f[(k-k_F)\xi_K]\ee
where the scaling function $f(y)$ obeys the normalization conditon:
\be \int dy f(y)=2\pi^2\ee
in order that $O(0)=1$.  Thus, 
\be O(R)=(1/2\pi^2)\int dy f(y)\sin [k_FR+(R/\xi_K)y]/k_FR.\ee
For $R\ll \xi_K$, this reduces to:
\be O(R)=\sin (k_FR)/k_FR\ee
independent of the details of the wave-function. Thus screening 
clouds centred on different impurities have negigible overlap 
provided that they are separated by a distance $R\gg k_F^{-1}$. 
A simple estimate of the condition for validity of the single 
impurity Kondo model is provided by the Noz\`ieres exhaustion principle. 
There must be enough available conduction electron states, 
within $\xi_K^{-1}$ of the Fermi wave-vector, to form 
linearly independent screening wave-functions around each impurity. 
This gives the conditions on the average impurity separation:
\bea n_{imp}^{-1/3}&\gg& \xi_K^{1/3}k_F^{-2/3}\ \  (3D)\\
 n_{imp}^{-1/2}&\gg& \xi_K^{1/2}k_F^{-1/2}\ \  (2D)\\
n_{imp}^{-1}&\gg& \xi_K\ \  (1D).\eea

\subsection{Knight Shift\label{sec:Knight}}
The Knight shift is proportional to the $r$-dependent polarization of 
the electron spin density in linear response to an applied magnetic field.
In NMR experiments, the total effective magnetic field felt by a nucleus 
is measured, from the nuclear resonance frequency. In addition to the 
applied magnetic field there is an additional contribution arising 
from the hyperfine interaction between the nuclear spin and the 
surrounding electron spins. Assuming for simplicity that this 
hyperfine interaction is very short range, the Knight-shift is just 
proportional to the local spin density at the location of the nucleus. 
For sufficiently weak applied fields, this is $\propto \chi (r)$, 
the local susceptibility. 
In the absence, of the magnetic impurity, this is simply the $r$-independent 
Pauli susceptibility, $\chi_0$. The Kondo interaction leads to an 
additional $r$-dependent term, which vanishes far from the impurity. 
Let
\be \vec S_{\hbox{el}}\equiv \int d^3rS_{\hbox{el}}(\vec r),\label{Sel}\ee
be the total electron spin operator.  The total conserved spin operator is:
\be \vec S_{\hbox{tot}}\equiv \vec S_{\hbox{imp}}+\vec S_{\hbox{el}}.
\label{Stot}\ee
 The $r$-dependent magnetic susceptibility
is:
\be \chi (r)\equiv <S^z_{\hbox{el}}(\vec r)S^z_{\hbox{tot}}>/T.\ee
[We ignore the possible difference of $g$-factors for impurity and 
conduction electrons for simplicity; this is discussed in Ref. 
~[\onlinecite{BA1,BA2}].] 
$\chi (r)$ contains a constant term, present at zero Kondo coupling, which is simply 
the usual Pauli susceptibility, $\nu /2$. 
Since only the s-wave harmonic couples to the impurity spin, the other terms 
come entirely from the s-wave component of the electron field. 
It is convenient to take advantange of the boundary condition of 
Eq. (\ref{freebc}), to make an ``unfolding transformation'', regarding 
$\psi_R(r)$ as the continuation of $\psi_L(r)$ to the negative $r$-axis:
\be \psi_R(r)=\psi_L(-r).\ee
The s-wave part of the spin density operator may then be written:
\be \vec S_{\hbox{el}}^{\hbox{s-wave}}(\vec r)\approx 
{1\over 8\pi r^2}[\vec J_L(r)+\vec J_L(-r)
+e^{2ik_Fr}\psi^\dagger_L(r){\vec \sigma \over 2}\psi_L(-r)++e^{-2ik_Fr
}\psi^\dagger_L(-r){\vec \sigma \over 2}\psi_L(r)]
\ee
We expect this  the continuum limit approximation
 to be valid at sufficiently long distances, $r\gg k_F^{-1}$. 
Thus, we see that the local susceptibility takes the form\cite{SA}
 at $r\gg k_F^{-1}$:
\be \chi (r)-\nu /2\to {1\over 8\pi^2r^2v_F}[\chi_{\hbox{un}}(r)+
\chi_{2k_F}(r)2\cos 2k_Fr ]\label{chired}\ee
where the uniform and $2k_F$ parts are slowly varying functions of $r$. 
These can be written in the 1D field theory as:
\bea \chi_{\hbox{un}}(r)&\equiv & v_F\int_0^\beta d\tau 
<[\psi^\dagger_L(r,\tau ){\sigma^z\over 2}\psi_L(r,\tau )+
\psi^\dagger_L(-r,\tau ){\sigma^z\over 2}\psi_L(-r,\tau )]
S^z_{\hbox{tot}}(0)>\nonumber \\
\chi_{2k_F}(r)&\equiv &- v_F\int_0^\beta d\tau 
<\psi^\dagger_L(r,\tau ){\sigma^z\over 2}\psi_L(-r,\tau )
S^z_{\hbox{tot}}(0)>.\label{cunkF}\eea 
Here $\vec S_{\hbox{tot}}$ is defined as in Eq. (\ref{Stot}) but 
we may restrict the electron part to its s-wave harmonic only, 
since the other harmonics give zero contribution to the 
Green's functions of Eq. (\ref{cunkF}):
\be \vec S_{\hbox{el}}={1\over 2\pi}\int_{-\infty}^\infty dr
\psi^\dagger_L(r){\vec \sigma \over 2}\psi_L(r)+\ldots .\ee

It can be proved,\cite{BA1,BA2} to all orders in perturbation theory, that 
\be \chi_{\hbox{un}}(r)=0.\ee
In the 1D field theory, 
$\chi_{2k_F}(r)$ obeys a simple renormalization group equation 
with zero anomalous dimension which expresses how 
physical quantities change under a change of ultraviolet 
cut off and of bare coupling:
\be \left[{\cal D}{\partial \over \partial {\cal D}}+\beta (\lambda ){\partial 
\over \partial \lambda }\right]\chi_{2k_F}(T,\lambda ,{\cal D},rT/v_F)=0.\ee
This simple form follows because only operators defined at $r=0$ have anomalous 
dimensions in a boundary field theory.  Thus there is no anomalous 
dimension for  $\psi^\dagger_L(r)\psi_L(-r)$. 
Furthermore, $\vec S_{\hbox{tot}}$ has zero anomalous dimension 
because it is conserved, commuting with the full Hamiltonian. 
This RG equation implies that $\chi_{2k_F}(T,\lambda_0 ,D,rT/v_F)$ 
which is, a priori a dimensionless function of 3 dimensionless 
variables, $\lambda_0$, (the bare coupling),
 ${\cal D}/T$ and $rT/v_F$, can be written as a 
function of the renormalized coupling at scale $T$, $\lambda (T)$ 
and $rT/v_F$ only. i.e. both the bare coupling, $\lambda_0$ and the 
ultraviolet cut-off, $D$ can be eliminated in favour of 
a single variable, $\lambda (T)$.  This is a basic consequence  
of renormalizability. An equivalent statement is that we may 
express $\chi_{2k_F}$ as a function of the ratio 
$rT/v_F$ and $\lambda (r)$, the 
effective coupling at energy scale $v_F/r$. 
This follows from the RG equation, Eq. (\ref{RGE}) 
which implies:
\be F[\lambda (r)]-F[\lambda (T)]=\ln (v_F/rT)\label{RG}\ee
where $F(\lambda )$ is the indefinite integral of $1/\beta (\lambda )$:
\be F(\lambda )\equiv \int^\lambda {1\over \beta (\lambda ')}d\lambda '.\ee
Eq. (\ref{RG}) can be solved to express $\lambda (T)$ in 
terms of $\lambda (r)$ and a function of $rT/v_F$. 

Results on $\chi_{2k_F}(r)$ were presented\cite{Gan,BA1,BA2} to third order in
 pertubation theory in the Kondo coupling, $\lambda_0$, at finite $T$. 
In the $T\to 0$ limit these become:
\be \chi_{2k_F}={\pi v_F\over 8rT}[\lambda_0+\lambda_0^2\ln (\tilde \Lambda r)
+\lambda_0^3\ln^2 (\tilde \Lambda r)+.5\lambda_0^3\ln (\tilde \Lambda r)
-\lambda_0^3\ln ({\cal D}/T)+\hbox{constant} \cdot \lambda_0^3].\label{chipert}\ee
Here $\tilde \Lambda \equiv 4\pi D/v_F$. 
The presence of the $\lambda_0^3\ln (D/T)$ term is very important. 
If this term were absent, we could safely take $T\to 0$ without 
obtaining any infrared divergence in perturbation theory, at 
sufficiently small $r$. In that case, it would be convenient 
to express $\chi_{2k_F}$ in terms of $\lambda (r)$ only.  
At small $\lambda (r)$ perturbation theory would apparently be valid. 
However, the presence of this term implies that this doesn't work. 
We can write instead:
\be \chi_{2k_F}\approx \pi {v_F\over 8rT}[\lambda (r)+\hbox{constant}\cdot 
\lambda^3(r)][1-\lambda (T)].\ee
An important lesson from this expression is that the behaviour 
of the local susceptibity does not become perturbative, at low $T$, 
even at small $r$. This contradicts the naive expectation from 
the analogy with QCD, discussed in Scc. I. 
 Nonetheless, it does suggest that there 
may be some crossover, or change in behaviour as we go from 
the region $r\ll \xi_K$ where $\lambda (r)\ll 1$ to 
$r \gg \xi_K$ where $\lambda (r)$ is large. A conjecture was 
made in Ref.~[\onlinecite{BA2}] for the behaviour at all $r\ll v_F/T$:
\be \chi_{2k_F}(r)\to {\pi v_F\over r}f(r/\xi_K)\chi (T),\label{conj}\ee
where $\chi (T)$ is the total, $r$-independent, impurity susceptility; 
i.e. the linear response of the total magnetization to an applied field, 
with the bulk Pauli term subtracted off. Here $f$ is some scaling 
function, depending on $r/\xi_K$. This conjecture is consistent 
with both the perturbative results given above, Fermi liquid 
results at long distances and low $T$, $r\gg \xi_K$, $T\ll T_K$, 
and also with results on the $k$-channel Kondo model in the large 
$k$ limit. $\chi (T)$ is well-known from 
a variety of methods, including BA. At $T\gg T_K$ it has the 
pertubative behaviour:
\be \chi (T)\to {1-\lambda (T)\over 4T}\approx {1\over 4T}
\left[1-{1\over \ln (T/T_K)}\right].\ee
At low $T\ll T_K$, $\chi (T)\to 1/(4T_K)$. For  
intermediate $T$ it is quite well approximated 
by $\chi (T)\approx .17/[(T+.65T_K)]$. We only know the 
form of the other scaling function, $f(r/\xi_K)$ asymptotically, 
at short and long distances.  
At short distances, $r\ll \xi_K$ it is perturbative:
\be f(r)\to \lambda (r)\approx {1\over \ln (\xi_K/r)}.\label{fshort}\ee
In the opposite limit, $r\gg \xi_K$ we may also determine\cite{SA}  
$f(r)$ from Fermi liquid theory. In the limit of low $T\ll T_K$ 
and large $r\gg \xi_K$ first order perturbation theory 
in the Fermi liquid interaction of Eq. (\ref{HFL}) gives the 
proposed form, eq. (\ref{conj}) with:
\be f(r/\xi_K)\to 4r/(\pi \xi_K) + 1/3 + O(\xi_K/r).\ee
At intermediate length scales, of $O(\xi_K)$, the function $f(r/\xi_K)$ 
must somehow crossover between $1/\ln (r/\xi_K)$ and $4r/(\pi \xi_K)$. 
Determining the function in this region requires numerical methods. 

At $T\gg T_K$, we found that the infrared divergences of perturbation theory 
are cut-off and weak coupling behaviour ensues, at any distance. 
There is no crossover at $r\approx \xi_K$ in this case. Our result to 
third order in $\lambda$ is:
\be \chi_{2k_F}(r)\approx (3\pi^2/4)\lambda (T)^2[1-\lambda (T)]e^{-2\pi rT/v_F}.\ee

Our proposed scaling form for the local susceptibility when $r\ll v_F/T$:
\be \chi (r) -\nu /2\to {\cos 2k_Fr\over 4\pi r^3}f(r/\xi_K)\chi (T)\label{chisum}\ee
makes it clear that NMR measurements of the Knight shift 
would be a very difficult way of detecting the Kondo screening cloud. 
Typically an NMR signal is only picked up for nuclei within 
a few times $k_F^{-1}$ of an impurity spin. Over this range, the $r$-dependence of Eq. 
(\ref{chisum}) is dominated by the first factor, $\cos 2k_Fr/r^3$. 
The slowly varying factor in the envelope of the oscillations, $f(r/\xi_K)$, 
could not be measured unless signals could be picked up out to distances 
of $O(\xi_K)$. At such large distances, the Knight shift 
is very small due to the factor of $1/r^3$. 
In fact, this form is qualitatively consistent with 
existing NMR experiments which looked for the Kondo screening 
cloud.\cite{Boyce} An important further complication in the experiments 
is that a given nucleus will feel a contribution to its Knight 
shift from many magnetic impurities at various distances.  
These contributions are expected to be simply additive, provided 
that the impurities are sufficiently diulte; see discussion at the 
beginning of Sec. \ref{sec:inf}.
Nonetheless, it would be extremely difficult to disentangle 
these various contributions and require extremely accurate measurements 
of the Knight shift to determined the tiny contributions from 
distance impurities. 

An important lesson here is that the Kondo screening cloud is 
difficult to detect because it is so large. To see it, one 
must look for a crossover at large distances but at these 
distances the effects are weak precisely because the distances 
are so large. 
The Kondo screening ``cloud'' is perhaps better described as a 
``faint fog''.\cite{Bergman} It is worth noting that 
the situation is much better for a sample of reduced dimensionality. 
Our formula for the local susceptibility, Eq. (\ref{chired}), 
 carries over directly to a 2D or 1D sample, with the factor of 
$1/r^2$ simply becoming proportional to $1/r^{D-1}$. In the 1D case, 
$\chi (r)/\cos (2k_Fr)$ crosses over from $1/[r\ln (\xi_K/r)]$ for $r\ll \xi_K$ 
to a constant at $r\gg \xi_K$, actually growing with distance. 
However, the diluteness condition is also modified in 1D to 
\be n_{\hbox{imp}}^{-1}\gg \xi_K\ee
as discussed at the beginning of Sec. \ref{sec:inf}.

\subsection{Density Oscillations\label{sec:Friedel}}
While the Kondo effect is associated with the spin degrees of freedom 
of the conduction electrons, it also has  an interesting effect on 
the charge density in the vicinity of the impurity.\cite{Mezei,ABS} This is perhaps 
a bit surprising, in light of spin-charge separation in 1D.  
It turns out that density oscillations only occur when particle-hole 
(p-h) symmetry is broken. In particular, none occur in a half-filled 
tight-binding model, which has exact particle-hole symmetry.  
Importantly, the Kondo interaction itself preserves p-h symmetry. 

These density oscillations turn out to be quite a bit simpler to 
analyse than the Knight shift, $\chi (r)$, discussed in sub-section \ref{sec:Knight}. 
This is because, at long distances, $r\gg k_F^{-1}$, they can be expressed as 
a half Fourier transform of a well-studied scaling function 
of $\omega /T_K$, the ${\cal T}$-matrix, ${\cal T}(\omega )$. Due to the $\delta$-function nature 
of the Kondo interaction, the exact retarded electron Green's function in the Kondo 
model can be written:
\be -i\int_0^\infty dt e^{i\omega t}<\{\psi(\vec r,t),\psi^\dagger (\vec r',0)\} >
=G(\vec r,\vec r',\omega )=G_0(\vec r-\vec r',\omega )
+G_0(\vec r,\omega ){\cal T}(\omega )G_0(-\vec r',\omega ).\ee
Here $G_0$ is the Green's function for the non-interacting case:
\be 
G_0(r,\omega )=\int {d^3k\over (2\pi )^3}e^{i\vec k\cdot \vec r}{1\over \omega -\epsilon_k
+i\eta }\label{G}\ee
The ${\cal T}$-matrix is used to determined the resistivity and thus plays 
an important role in the theory of the Kondo model.  Although it is not 
accessible by BA, NRG techniques have been developed to calculate it, 
in addition to perturbative and Fermi liquid results and other approximations. 
The density is obtained from the Green's function in the standard way:
\be \rho (r)=-{2\over \pi}\int_{-\infty}^0d\omega \hbox{Im}
G(\vec r,\vec r,\omega).\label{rhoG}\ee
Note that $G(\vec r,\vec r,\omega )$ itself has trivial $r$-dependence, determined 
entirely by $G_0(r)$.  Only the frequency-dependence reflects the Kondo physics. 
However, once we integrate over $\omega$ in Eq. (\ref{rhoG}), this $\omega$-dependence 
introduces some interesting $r$-dependence. In the limit $r\gg k_F^{-1}$, we 
may use the asymptotic expression for $G_0(r)^2$.  In D-dimensions, for the free-electron 
dispersion relation this is:
\be G_0(r,\omega )^2\to -{1\over v_F^2}\left({-ik_F\over 2\pi r}\right)^{D-1}\exp (2ik_Fr
+2i\omega r/v_F).\label{G_0}\ee
This asymptotic form of $G_0$ can be calculated for other dispersion relations 
and has a similar form. Thus we see that, at $r\gg k_F^{-1}$ the density 
oscillations induced by a Kondo impurity are given by the half-Fourier transform 
of the ${\cal T}$-matrix. Furthermore, for such large $r$, we expect this 
half Fourier transform to be dominated by frequencies $\leq v_F/r \ll v_Fk_F \propto D$. 
In this frequency range, the ${\cal T}$-matrix is a universal scaling function of 
$\omega /T_K$. 

It is important in calculating the density oscillations to take into account additional 
potential scattering, arising from the magnetic impurity, in addition to the Kondo interaction. 
This was ignored in sub-section \ref{sec:Knight} since it is not very important for the Knight shift. 
This corresponds to an additional term in the Kondo Hamiltonian, Eq. (\ref{HK}) of the form:
\be \delta H=V\sum_{\sigma}\psi^\dagger_\sigma (0)\psi_\sigma (0).\ee
We refer to the s-wave phase shift at the Fermi surface due to the potential scattering 
as $\delta_P$. The ${\cal T}$-matrix, at $\omega \ll D$ can be expressed in terms 
of $\delta_P$, a universal function, $t_K(\omega /T_K)$ and the density 
of states in D-dimensions, $\nu_D$:
\be {\cal T}(\omega )=\left\{ e^{2i\delta_P}[t_K(\omega /T_K)+i]-i\right\}/(2\pi \nu_D).
\label{Tt}\ee
Note that, in the particle-hole symmetric case, $\delta_P=0$, ${\cal T}=t_K/(2\pi \nu_D)$. 
By rescaling the frequency integration variable, and using $v_F/T_K=\xi_K$,
  we see that the half-Fourier transform 
occuring in $\rho (r)$ may be expressed in terms of a universal scaling function, 
$F(r/\xi_K)$:
\be \int_{-\infty}^0d\omega e^{2i\omega r/v_F}t_K(\omega /T_K)\equiv [v_F/(2r)][F(rT_K/v_F)-1].
\label{Fdef}\ee
Particle-hole symmetry of the model with no potential scattering (from which $t_K$ can 
be determined) implies that $t_K^*(\omega /T_K)=-t_K(-\omega /T_K)$. Furthermore, 
since $t_K$ is related to the retarded Green's function it is analytic in the upper half plane. 
These two conditions imply that $F(r/\xi_K)$ is real. On the other hand, note 
that it depends on both real and imaginary parts of $t_K$. This makes it challenging 
to determine the density oscillations numerically, since while reasonably accurate results may exist 
for Im $t_K$, this is not the case for the real part. From Eqs. (\ref{rhoG}), (\ref{Tt}) 
and (\ref{Fdef}) we obtain the formula for the density oscillations at long 
distances from the impurity, $r\gg k_F^{-1}$:
\be \rho (r)-\rho_0\to {C_D\over r^D}[\cos (2k_Fr-\pi D/2+2\delta_P)F(r/\xi_K)
-\cos (2k_Fr-\pi D/2)]\label{friedel}\ee
where the constant is $C_3=1/(4\pi^2)$, $C_2=1/(2\pi^2)$ and $C_1=1/(2\pi )$. 
The formula applies in D=1,2 or 3, with a spherically symmetric dispersion 
relation.  It also applies to the 1D tight-binding 
model in which case $r$ is restricted to integer values. Note that in this case, 
there is exact p-h symmetry at half-filling with no potential scattering, 
corresponding to $k_F=\pi /2$ and $\delta_P=0$. In this case, the right 
hand side of Eq. (\ref{friedel}) vanishes identically as expected. 
p-h symmetry breaking is necessary to get non-zero density oscillations 
and the potential scattering phase shift, $\delta_P$ plays an important 
role in them. 

The universal Kondo ${\cal T}$-matrix, $t_K(\omega /T_K)$, can be calculated 
at higher frequencies, $\omega \gg T_K$ using  perturbation theory 
in the Kondo interaction.  It can be calculated at low frequencies using 
Fermi liquid theory. We find that these expansions can be simply inserted 
into Eq. (\ref{Fdef}) to obtain expansions for $F(r/\xi_K)$ and hence 
the density oscillations at short distances, $r\ll \xi_K$ and long 
distances, $r\gg \xi_K$. The weak coupling result is
\be t_K(\omega )=-(3i\pi^2/8)[\lambda_0^2+2\lambda_0^3\ln ({\cal D}/\omega )+\ldots ].
\label{texp}\ee
This may be written:
\be t_K(\omega /T_K)\to -(3\pi^2i/8)\lambda^2 (|\omega |)\approx -3\pi^2i/[8\ln^2(|\omega |/T_K)].\ee
The ${\cal T}$-matrix vanishes slowly at high frequencies. Substituting this weak 
coupling expansion into the integral defining $F$, Eq. (\ref{Fdef}) and integrating 
term by term gives:
\be F=1-(3\pi^2/8)[\lambda_0^2+2\lambda_0^3\ln (\tilde \Lambda r)+\ldots ]\ee
where $\tilde \Lambda$ is of order $v_F/D$. We recognize the weak couplig expansion 
of the effective coupling at scale $r$:
\be F(r/\xi_K)\to 1-(3\pi^2/8)\lambda^2 (r)\approx 1-3\pi^2/[8\ln^2(\xi_K/r)].\ee
Note that in the weak coupling limit $F\to 1$ and then Eq. (\ref{friedel}) 
reduces to the standard formula for the Friedel oscillations induced by 
a non-magnetic impurity which produces a phase shift at the Fermi surface of $\delta_P$ 
(in the s-wave channel only). 

Our results indicate that weak coupling behaviour occurs for the density oscillations 
at short distances $r\ll \xi_K$ even at $T=0$. This is quite different than 
what we found for the Knight shift in Sec. \ref{sec:Knight}. In that case we found, in Eq. (\ref{chipert}),  
that  the infrared divergences of perturbation theory 
were not completely cut off at short distances.  However our conjecture, Eq. (\ref{conj}), 
implied that the non-perturbative aspect simply led, at low $T\ll v_F/r$, to a factor of $\chi (T)$ 
in $\chi (r)$ which was otherwise linear in $\lambda (r)$, Eq. (\ref{fshort}). 
The density oscillations therefore show simpler behaviour at short distances. 

At low frequencies, $\omega \ll T_K$, the Kondo ${\cal T}$-matrix can be calculated from 
Fermi liquid theory:
\be t_K\to -i[2+i\pi \omega /(2 T_K)-3\pi^2\omega^2/(16T_K^2)+\ldots ].\ee
Substituting into Eq. (\ref{Fdef}), gives:
\be F(r/\xi_K)\to -1+\pi \xi_K/(4r)-3\pi^2\xi_K^2/(32r^2)+\ldots \ee
At large distances, $r\gg \xi_K$, $F\to -1$. From Eq. (\ref{friedel}), we 
see that this is equivalent to shifting $\delta_P\to \delta_P+\pi /2$ 
while setting $F=1$, 
the non-interacting value. 
In this limit the Kondo scattering simply changes the phase shift 
by  $\pi /2$, the familiar result. Thus, while we obtain the standard Friedel oscillation 
formula at both short and long distances, there is a non-trivial and universal crossover, 
determined by $F(r/\xi_K)$, at intermediate distances. 

\begin{figure}[!ht]
\begin{center}
\includegraphics[height=7cm]{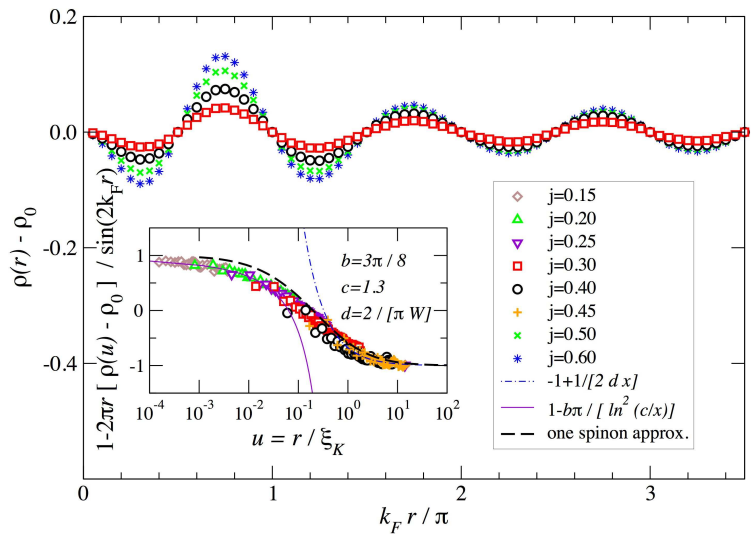}
\end{center}
\caption{NRG results on charge oscillations around a Kondo impurity
coupled to 1D conduction electrons with particle-hole symmetry  from Ref. [\onlinecite{ABS}].
 Note that the oscillations vanish at
$k_Fr/\pi\in\mathbb{N}$. As shown in the inset,
the properly rescaled envelope function of the oscillations  
(extracted as $\rho-\rho_0$ at the local maxima)
for different Kondo couplings
nicely collapse into one universal curve except for the points
where $r\sim k_F^{-1}$. 
In the inset we show the analytical results
for the asymptotics as well: Note the good agreement between the analytical
results and the numerics.}
\label{fig:friedel}
\end{figure}

The density oscillations were calculated using an improved NRG method in Ref.~[\onlinecite{ABS}]. 
In this approach Wannier states are introduced both at the impurity location and at the 
point of interest, $r$, fixing the problem with spatial resolution 
in the usual NRG approach.  Results are shown in Fig. (\ref{fig:friedel}).  The oscillations 
can be parametrized, at $r\gg k_F^{-1}$, by the form of Eq. (\ref{friedel}).  The function 
$F(r/\xi_K)$ fits the asymptotic predictions at large and small arguments, crossing 
over between $1$ and $-1$ as expected. In fact, it is fairly well fit, throughout the crossover 
by the simple ``one spinon approximation''\cite{Mezei,Lesage}
\be t_K(\omega /T_K)\approx -2i/[1-i\pi /(2T_K)]\ee
giving an approximate formula for $F$:
\be F(u)\approx 1+(8u/\pi )e^{4u/\pi}\hbox{Ei}(-2u/\pi ),\ee
where $E_i(x)$ is the exponential-integral function.

Experimental measurement of the Kondo screening cloud via density oscillations 
would again be extremely challenging. The $1/r^D$ factor in Eq. (\ref{friedel}) 
implies that the oscillations would be extremely small at the length scale $\xi_K$. 
(Note that the number of oscillations that would need to be measured is of 
order $\xi_Kk_F\gg 1$.) As for the Knight shift, the situation is improved 
in lower dimensions. The Kondo effect is apparently observed for 
magnetic impurities on metallic surfaces using Scanning Tunelling Microscopy (STM). 
In cases, where the Kondo interaction is predominantly with surface conduction 
electron states, Eq. (\ref{friedel}) with $D=2$ should apply. The STM 
tunelling rate for tip energy $E$ is usually assumed to be proportional 
Eq. (\ref{rhoG}) but with the lower limit of integration replaced by $E$. 
This quantity could also be calculated from knowledge of $t_K$, but only 
when $E\gg T_K$ do we recover a scaling function of $r/\xi_K$ only.

\subsection{Impurity spin correlation function\label{sec:corr}}
Another quantity which gives a very direct picture of the Kondo 
screening cloud is the equal time ground state correlator of the 
impurity spin with the spin density at a distance $r$:
\be K(\vec r,T)\equiv <0|S^z_{\hbox{el}}(\vec r,0)S^z_{\hbox{imp}}(0)|0)>.\ee
Although this quantity was also discussed at finite $T$ in Ref.~[\onlinecite{BA1,BA2}], 
we restrict our discussion here to the 
$T=0$ case. Unlike for the local susceptibility, no divergences 
are encountered in perturbation theory by taking $T\to 0$, 
at least to the (third) order studied. 
$K(r)$ can easily be seen to obey an exact sum rule.  If we consider 
a finite system with an even number of electrons in total 
(including the impurity spin as an electron) then the ground state 
is a spin singlet:
\be S_{\hbox{tot}}(0)^z|0>=0.\ee
Thus we have:
\be 0=<0| S^z_{\hbox{imp}}(0)S_{\hbox{tot}}(0)^z|0>=1/4+\int d^3rK(\vec r)
.\label{rule}\ee
A heuristic picture of $K(r)$ is obtained by writing:
\be S^z_{\hbox{el}}(\vec r)\approx s^z_{\hbox{el}}\rho (\vec r)\ee
Here $\rho (\vec r)$ is the probability of the screening electron 
being at the location $\vec r$ and $s^z_{\hbox{el}}$ is a S=1/2
operator, representing the spin of the screening electon, with
$<S^z_{\hbox{imp}}s^z_{\hbox{el}}>=-1/4$. 
$\rho (\vec r)=|\phi (\vec r)|^2$ 
where $\phi (\vec r)$ is the screening cloud wave-function. 
Then we obtain $K(r)\approx -\rho (r)/4$ and 
the sum rule of Eq. (\ref{rule}) is simply the normalization 
condition on the screening electron's wave-function. 
As for the local susceptibility, discussed in sub-section \ref{sec:Knight}, 
at $r\gg k_F^{-1}$ we may decompose $K$ into uniform and $2k_F$ parts:
\be K(r)\to {1\over 8\pi^2v_Fr^2}[K_{\hbox{un}}(r)+2K_{2k_F}(r)]\ee
where $K_{\hbox{un}}$ and $K_{2k_F}$ can be calculated in the 
1D field theory. Perturbation theory in the Kondo coupling, 
valid at distances $r\ll \xi_K$,  gives:\cite{BA1,BA2}
\bea K_{\hbox{un}}&\approx & -{\pi v_F\lambda (r)^2(1+\lambda_0/2)\over 2r}
\nonumber \\
K_{2k_F}&\approx & {\pi v_F\lambda (r)(1+\lambda_0/2)\over 8r}
\eea
Several differences are to be noted with the local susceptibility. 
First of all, we have obtained a finite result at $T=0$, as noted 
above. Secondly the uniform part is now non-zero.  Thirdly, 
the result cannot be expressed in terms of the renormalized 
coupling at scale $r$ only, but also involves a factor 
containing the bare coupling, $\lambda_0$. This can be understood\cite{BA1,BA2} 
from the fact that the impurity spin operator has a non-zero anomalous 
dimension:
\be \gamma_{\hbox{imp}}\approx \lambda^2/2+\ldots \ee
However, this is not very important since this factor goes to 1 
in the scaling  limit of weak bare coupling. 
On the other hand, at $r\gg \xi_K$ we may calculate\cite{BA1,BA2,Ishii}
 $K(r)$ using Fermi liquid theory giving:
\be K_{2k_F}\to -(1/2)K_{\hbox{un}}\to \hbox{const}\times {\xi_Kv_F\over r^2}.\ee
Our heuristic interpretation then suggests a slow power-law decay 
for the  the screening cloud electron probability density:
\be \rho (r)\approx \hbox{const}\times {\sin^2k_Fr \xi_K\over r^4}
\ \ (r\gg \xi_K).\ee
On the other hand, at shorter distances, $r\ll \xi_K$, $K(r)$ is not 
negative definite and the heuristic picture of 
$K(r)$ is not valid.

\section{Spin in a mesoscopic device\label{sec:meso}}
Given the extreme difficulty of observing the Kondo screening cloud
from length-dependent measurements in a macroscopic sample,
discussed in Sec. \ref{sec:inf},  together 
with the fact that mesoscopic devices routinely contain 
components with dimensions of order 1 micron, it is natural 
to consider whether the best way of finally observing the Kondo 
screening cloud experimentally might be via size-dependent effects 
in such a device. We are interested in devices containing a quantum 
dot in the Coulomb blockade regime with a gate voltage tuned 
so that the number of electrons in the dot is odd, with 
a S=1/2 ground state. For simplicity, we will also assume 
that the tunelling from dot to leads is sufficiently weak 
that only virtual charge fluctuations of the dot need be considered 
so that the Kondo model is appropriate, rather than the Anderson model. 
Alternatively, we might  consider one or more magnetic impurity 
atoms in a mesoscopic sample. 

An important point is that, associated with a finite sample 
size, we have a finite spacing between energy levels.  
If the sample is 1 dimensional, then the level spacing 
will generally be $\Delta \approx v_F/L$ where $L$ is the sample size. 
Thus the condition $v_F/T_K\approx L$ is equivalent to $\Delta\approx T_K$. 
Thus, it sometimes argued that observing size dependence in this 
situation does not really show the existence of a Kondo screening cloud 
but ``merely'' that $T_K$ is the characteristic energy scale for the Kondo 
model. However, since as discussed in the previous sections, the Kondo 
screening cloud concept is really just the inevitable consequence 
of being able to convert an energy scale to a length scale using 
a factor of velocity, size effects can provide realizations of it. 
We may equally well say that the behaviour changes when the 
finite size gap becomes larger than $T_K$ or else when 
the Kondo screening cloud no longer fits inside the sample. 
If the sample should instead be regarded as 2 or 3 dimensional 
the finite size level spacing may be $\ll v_F/L$ depending 
on which energy levels are important and other details, discussed in 
Sec. \ref{sec:box}. 

The Kondo model is used both to describe magnetic impurities in 
metals and also gated semi-conductor heterostructure quantum dots. 
The heterostructure (such as GaAs/AlGaAs) provides a fairly clean 2D electron gas (2DEG) 
buried some distance (often around 100 Angstroms) below the surface 
of the semi-conductor wafer. Gate voltages are applied to the surface to 
define point contacts and quantum dots. A quantum dot refers to 
an island of electrons, with the electron number (typically around 100 or less) 
controlled in unit steps by a gate voltage. When this number is odd, 
the quantum dot usually has an S=1/2 ground state. The quantum dot 
can be connected by narrow point contacts to the left and right sides 
of the 2DEG. If the point contacts are close to being pinched off 
they only permit one channel of electrons to pass through, giving 
a $2e^2/h$ conductance. In this case, a simplified model of the system 
involves effectively 1D leads.   The quantum dot has a charging energy, $U$, 
as well as a gate voltage, such that the energy as a function of 
electron number, $N$, is $U(N-N_0)^2$ for a value of the parameter $N_0$
controlled by the gate.  It is often convenient, theoretically, 
to use a 1D tight-binding model; universality of Kondo physics implies 
that such details are not important. The corresponding Hamiltonian is:
\be H=H_0+H_d\label{HAnd}\ee
where
\be H_0=-\left[ t\sum_{j=-\infty}^{-2}c^\dagger_{j\alpha}c_{j+1,\alpha}
-t\sum_{j=1}^{\infty}c^\dagger_{j\alpha}c_{j+1,\alpha}-t'c^\dagger_{-1\alpha}c_{0\alpha}
-t'c^\dagger_{0\alpha}c_{1\alpha}+h.c.\right]\label{H0}\ee
and
 \be H_d=\epsilon_0n_0+Un_{0\uparrow}n_{0\downarrow}.\ee
Here $n_{0\alpha}$ is the occupation number at site $0$ for spin $\alpha$, 
with $n_0\equiv n_{0\uparrow}+n_{0\downarrow}$. $t'$ represents 
the tunelling amplitudes, through the left and right point 
contacts, to the leads. (We take these to be equal for simplicity 
but the generalization is straightforward.)

If the tunnelling amplitudes are sufficiently weak 
compared to $U$, then the quantum dot makes only virtual charge fluctuations 
and the system is in the Coulomb blockade regime.  The conductance 
through the dot then usually begins to decrease as the temperature is 
lowered, if $N_0$ is close to an integer. In the case where $N_0$ is near an {\it odd} 
integer, so that the quantum dot behaves as an S=1/2 impurity, 
we may  make a Shrieffer-Wolff transformation to the corresponding Kondo 
model with Hamitonian:
\be H=H_0+H_K\label{HKemb}\ee
where now:
\be H_0=-\left[ t\sum_{j=-\infty}^{-2}c^\dagger_{j\alpha}c_{j+1,\alpha}
-t\sum_{j=1}^{\infty}c^\dagger_{j\alpha}c_{j+1,\alpha} \right]
\label{H0emb}\ee
\be H_K=J(c_{-1}+c_1)^\dagger{\sigma \over 2}(c_1+c_{-1})\cdot \vec S\label{Hintemb}\ee
where $\vec S$ is the spin operator for the site $0$:
\be S\equiv c^\dagger_0{\vec \sigma \over 2}c_0\ee
and the Kondo coupling is:
\be J=2t'^2\left[{1\over -\epsilon_0}+{1\over U+\epsilon_0}\right].\ee
In general, a potential scattering term is also generated, of 
the same order of magnitude as $J$. To simplify the 
discussion, we focus here on the particle-hole symmetric case, 
$\epsilon_0=-U/2$, where the potential scattering vanishes. 
We also assume that the system is at half-filling so that 
the particle-hole symmetry is exact.  It is then easy to 
calculate the zero temperature conductance for this Kondo 
model in either limit, $J\ll t$ or $J\gg t$. The 
second limit is not physical for the underlying Anderson model 
of Eq. (\ref{HAnd}) but it is convenient to consider 
nonetheless as a low energy fixed point Hamiltonian. 
When $J=0$ the the Kondo Hamiltonian contains no terms linking 
left and right sides of the dot, so the conductance vanishes. 
When $J\gg t$, one 
electron gets trapped in the symmetric orbital, with annihilation operator:
\be c_s\equiv (c_1 +c_{-1})/\sqrt{2}\ee
 and forms 
a singlet with the impurity spin. At first sight, one 
might think that this would block the transport through 
the impurity. However, this is not so due to transport 
through the antisymmetric orbital with annihilation operator:
\be c_a\equiv (c_1-c_{-1})/\sqrt{2}.\ee
Electrons entering from sites $-2$ or $2$ can hop into this 
antisymmetric orbital without breaking the Kondo singlet. If instead 
they hop into the symmetric orbital there is a large energy cost 
of $O(J)$. 
At large $J/t$ we may obtain a low energy effective Hamiltonian 
from $H_0$ by projecting $c_{\pm 1}$ onto $c_a$:
\be Pc_{\pm 1}P=\pm c_a/\sqrt{2}.\ee
This gives:
\be H_{\hbox{low}}=-t\left[ \sum_{j=-\infty}^{-3}c^\dagger_{j\alpha}c_{j+1,\alpha}
-\sum_{j=4}^{\infty}c^\dagger_{j\alpha}c_{j+1,\alpha}+{1\over \sqrt{2}}c^\dagger_{-2\alpha}c_{a\alpha}
-{1\over \sqrt{2}}c^\dagger_{2\alpha}c_{a\alpha}
+h.c.\right]\label{Hlow}\ee
The tranmission probability for this non-interacting electron problem 
is easily calculated:\cite{Simon1}
\be T(k)=\sin^2k.\ee
In particular, at half-filling, $k_F=\pi /2$, the transmission probability 
at the Fermi energy is one (corresponding to a resonance). The conductance for this non-interacting 
model is given by the Landauer formula:
\be G={2e^2\over h}T(\epsilon_F)={2e^2\over h}.\ee
Calculations show that, 
for a small bare Kondo coupling,  the 
conductance  grows with decreasing $T$, saturating at the 
perfect tranmission value, $2e^2/h$, for $T\ll T_K$. 
These results can be obtained by expressing the conductance 
as a frequency integral of the imaginary part of the ${\cal T}$-matrix, 
introduced in sub-section \ref{sec:Friedel}, at finite temperature, times 
the derivative of the Fermi function. Analytic formulas 
can be obtained for $G(T)$ in both the high $T$ and low $T$ limits 
using perturbation theory and Fermi liquid theory respectively. 
Calculations at intermediate $T$ are typically  based on 
less controlled approximations. This increase of conductance
with decreasing $T$, due to the Kondo effect, only 
sets in at low temperatures, following an initial decrease due 
to the onset of the Coulomb blockade.

Note that this behaviour is the inverse of what happens for 
a magnetic impurity in a bulk metal where it is the resistivity 
which grows with lowering $T$, not the conductance. The impurity behavior
is more closely related to that for a side-coupled quantum dot where 
the conductance is $2e^2/h$ for zero Kondo coupling (high $T$) 
and vanishes at low $T\ll T_K$. 

\subsection{Persistent current in a ring containing a Kondo impurity\label{sec:pers}}
\begin{figure}[!ht]
\begin{center}
\includegraphics[height=7cm]{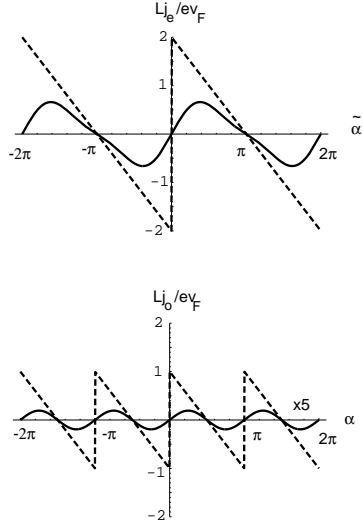}
\end{center}
\caption{Persistent current vs. flux for an even or odd number of 
electrons in the weak coupling limit, for $\xi_K/L\approx 50$, from Eq. (\ref{jpert})
(solid line) and in the strong coupling limit, $\xi_K/L<<1$, from Eq. (\ref{jstrong}) 
 (dashed line).  $j_o$ 
is multiplied $\times 5$ at $\xi_K/L=50$ for visibility.  The 
solid lines are obtained from Eq. (\ref{jpert}) using the effective 
coupling $\lambda (L) \approx 1/\ln (\xi_K/L)$. (From~[\onlinecite{AS}].)}
\label{fig:PCpert}
\end{figure}

\begin{figure}[!ht]
\begin{center}
\includegraphics[height=7cm]{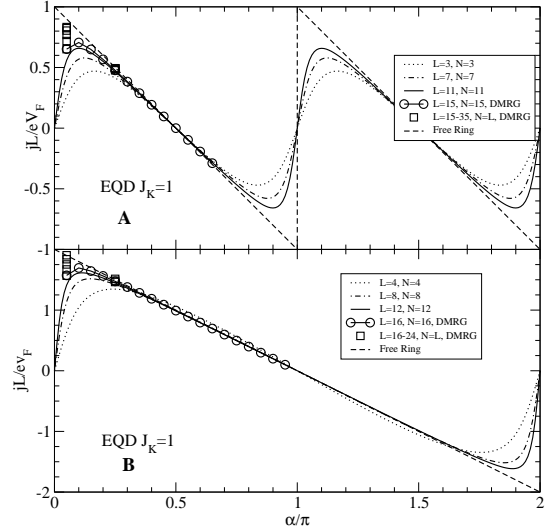}
\end{center}
\caption{The current at 1/2-filling for $N=4p-1$ (A) and $N=4p$ (B) for the embedded quantum dot.
 Results\cite{SA2}
are shown for a number of system sizes with $J_K=1$ as a function of $\alpha/\pi$. 
The results for the small system sizes
are obtained using exact diagonalization methods while the larger system sizes (circles) have been
obtained using DMRG techniques. From Ref. [\onlinecite{SA2}].}
\label{fig:PC}
\end{figure}

A calculationally simple situation in which to observe finite size effects 
is to close the leads embedding the quantum dot into a ring. The Hamiltonian 
is that of Eqs. (\ref{HKemb}-\ref{Hintemb}) but the 
leads are now of finite length with periodic boundary conditions:
\bea H_0&=&-t\sum_{j=1}^{L-2}(c^\dagger_{j\alpha}c_{j+1,\alpha}+h.c.)\label{H0ring}\\
H_K&=&J(c_{L-1}+c_1)^\dagger {\vec \sigma \over 2}(c_{L-1}+c_1)\cdot \vec S.\eea
 While the 
conductance is now not measurable, one can instead calculate the 
persistent current in response to an enclosed magnetic flux, $\Phi = (\hbar c/e)\alpha$.
 The current, at $T=0$, is given by the derivative of the ground state energy 
with respect to the flux:
\be j=-(e/\hbar )dE_0/d\alpha .\ee
A perturbative calculation gives:\cite{AS,Simon1}
 \begin{eqnarray}
j_e(\alpha ) &=& {3\pi v_F e\over 4  L}\{ [\sin \tilde \alpha
[\lambda + \lambda^2\ln (Lc)] +(1/4 +\ln 2)\lambda^2\sin
2\tilde \alpha \} + O(\lambda^3)\ \  \nonumber \\
j_o(\alpha )&=&{3\pi v_F e\over 16  L}\sin 2\alpha
[\lambda^2+2\lambda^3\ln (Lc')]+O(\lambda^4),\label{jpert}\end{eqnarray}
for $N$ even and odd respectively,  where $c$ and $c'$ are
constants of O(1) which we have not determined and:
\begin{eqnarray}
\tilde \alpha &=& \alpha \ \  (N/2\  \hbox{even})\nonumber \\
\tilde \alpha &=& \alpha + \pi \  \  (N/2\
\hbox{odd}).\label{tildedef}\end{eqnarray}
$N$ is the number of electrons, including the electron on the quantum 
dot, $j=0$. At half-filling, $N$ is just the total number of lattice 
sites, including the origin; $N=L$. The fact that $j$ is O($\lambda )$ 
for $N$ even but not for $N$ odd is easily understood. The unperturbed 
ground state consists of a partially filled Fermi sea and a decoupled 
impurity spin. For $N$ even, there are an odd number, $N-1$, of 
electrons in the Fermi sea.  The unpaired electron at the Fermi 
surface forms a spin singlet with the impurity in first 
order degenerate perturbation theory. On the other hand, for $N$ 
odd, there are no unpaired electrons in the non-interacting Fermi 
sea so it is necessary to go to second order in $\lambda$. 
A very important property of these results is that, to the order worked, 
the persistent current only depends on the effective Kondo coupling 
at the length scale $L$: $\lambda (L)=\lambda+\lambda^2\ln L$. That is, logarithmic divergences are encountered 
in next to leading order, as is standard for many calculations in the 
Kondo model, but they only involve $\ln L$, not $\ln T$.  The current 
is finite at $T=0$, for finite $L$. This suggests that the finite 
size of the ring is acting as an infrared cut-off on the growth 
of the Kondo coupling.  Provided that $\lambda \ln L \ll 1$, the 
higher order corrections are relatively small and perturbation theory 
appears trustworthy. This is equivalent to the condition $\xi_K\gg L$. 
We may say that, in this case, the screening cloud doesn't ``fit''
inside the ring so the Kondo effect (growth of the effective 
coupling to large values) doesn't occur. We expect that 
higher order perturbation theory would preserve this property, 
giving $v_Fe/L$ times functions of the renormalized coupling 
at scale $L$ (and the flux) only. This follows because the 
current obeys a renormalization group equation with zero 
anomalous dimension. This in turn follows from the fact that the 
current is conserved $d<j(x)>/dx=0$ so it can be calculated 
at a point far from the impurity spin. This scaling form 
implies, equivalently, that we can write the current as:
\be j_{e/o}={ev_F\over L}f_{e/o}(L/\xi_K,\tilde \alpha )\label{PCscale}\ee
for even and odd $N$ respectively, 
where $f_e$ and $f_o$ are universal scaling functions, depending only 
on the ratio $L/\xi_K$ and not separately on the bare 
Kondo coupling $\lambda_0$ and cut-off ${\cal D}$.  In 
the limit $L\gg \xi_K$, the persistent current can be calculated 
using the non-interacting effective low energy Hamiltonian 
of Eq. (\ref{Hlow}), i.e. Fermi liquid theory.  This gives:
\bea j_{e}(\alpha )&=& -{2ev_F\over \pi L}[\tilde \alpha -\pi ]\ \  (N\ \hbox{even}),\\
j_{o}(\alpha )&=& -{ev_F\over \pi L}([\alpha ]+[\pi -\pi ]) ]\ \  (N\ \hbox{odd}).
\label{jstrong}\eea
Here 
\be [\alpha ]\equiv \alpha  \ (\hbox{mod}\ 2\pi ),\ \  |[\alpha ]|\leq \pi .\ee
These small $L$ and large $L$ limits are plotted in Fig. (\ref{fig:PCpert}). 
Note that the current has the same sign for all $\alpha$ in both 
limits.  Furthermore it has the same period: $2\pi$ for $N$ even 
and $\pi$ for $N$ odd. For intermediate lengths, $L$ of order $\xi_K$, 
it is necessary to do a numerical calculation. A combination of exact diagonalization 
and DMRG~[\onlinecite{SA2}] supports the scaling behavior of Eq. (\ref{PCscale}) giving scaling 
functions that interpolate smoothly, as a function of $\xi_K/L$.  See. Fig. (\ref{fig:PC}). 

\subsection{Charge steps in a finite length quantum wire terminated by a 
quantum dot\label{sec:steps}}
\begin{figure}[!ht]
\begin{center}
\includegraphics[height=4cm]{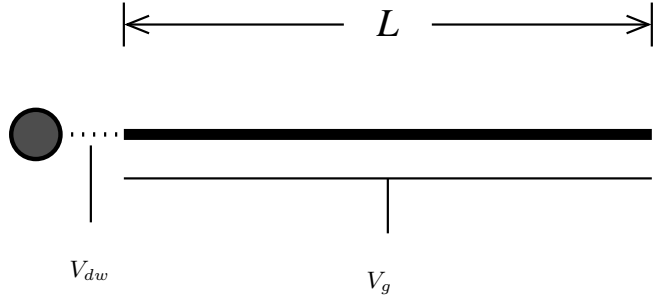}
\end{center}
\caption{Possible experimental setup. $V_{dw}$ controls the tunneling $t^{\prime}$ between
the small dot (on the left) and the wire and $V_{g}$ varies the chemical
potential in the wire.}
\label{fig:expsetup}
\end{figure}

\begin{figure}[!ht]
\begin{center}
\includegraphics[height=7cm]{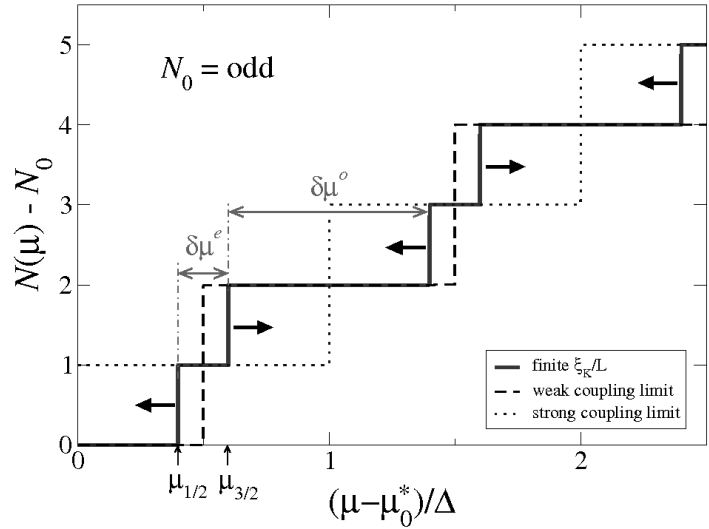}
\end{center}
\caption{ Charge quantization steps for the wire coupled to
a small quantum dot. The arrows indicate the direction in which the
single steps move as $\lambda\left(L\right)$ grows.}
\label{fig:steps}
\end{figure}

Another mesoscopic system which is quite readily analysed theoretically involves 
a linear quantum wire of finite length with a quatum dot, in the Kondo regime, at one end. 
[See Fig. (\ref{fig:expsetup}).]
We assume that the system is very weakly tunnel-coupled to an electron reservoir and that 
a uniform gate voltage, $V_G$, is applied to the wire, corresponding 
to a chemical potential $\mu =-eV_G$, in addition to the gates controlling the 
occupancy of the quantum dot and the tunelling between dot and wire. As $\mu$ is varied, 
the number of electrons in the system, in its ground state will vary in single steps 
at particular values of $\mu$. This defines a ``charge staircase''. See Fig. (\ref{fig:steps}).
The width of 
the steps, with an even or odd number of electrons in the system, is a sensitive 
indicator of the strength of the effective Kondo coupling. The steps, at discrete values of 
$\mu$, could be detected by some Coulomb blockade technique. 
Such measurements are likely to be far easier than measuring persistent currents in closed rings. 
For simplicity, we treat 
the wire as having only a single channel - i.e. as being an ideal 1D system. A great 
advantage of this system is that the charge staircase can be determined very accurately 
from the Bethe ansatz solution~[\onlinecite{Andrei,Weigmann}] of the Kondo model. Remarkably, 
this remains true, to some extent, even when we include short range Coulomb interactions 
throughout the wire. In this case we take advantage of the more recent Bethe ansatz solution 
of an S=1/2 Heisenberg antiferromagnet with a weakly coupled spin at one end~[\onlinecite{Frahm}]. 
We first discuss the case with no Coulomb interactions (apart from those in the quantum dot, 
leading to the effective Kondo model) and later include bulk Coulomb interactions in the wire.

\psfrag{Vg}{$V_g$}
\psfrag{Vdw}{$V_{dw}$}

\subsubsection{Interactions in quantum dot only\label{sec:noCoulomb}}
Again it is convenient to consider a tight-binding model.  In the Kondo limit this is 
\be H=H_0+H_K\ee
with 
\be H_0=-t\sum_{j=1}^{L-2}[(c^\dagger_{j\alpha}c_{j+1,\alpha}+h.c.)-\mu c^\dagger_jc_j]
\label{H0QW}\ee
and
\be H_K=Jc_1^\dagger {\vec \sigma \over 2}c_1\cdot \vec S.\label{HKQW}\ee
Note that this time we have ``open'' boundary conditions. The chain terminates at site $L-1$. 
We are interested in how the total electron number in the system, $N$, changes at $T=0$ in the
 grand canonical 
ensemble, as we vary $\mu$.  We define $N$ to include the single electron on the quantum dot. 
Considering a small range of $\mu$, the steps with even $N$ all have the same width, 
as do the steps with odd $N$.  We are interested in the ratio $R$:
\be R\equiv {\delta \mu^o\over \delta \mu^e}\ee
where $\delta \mu^{e/o}$ is the width of the interval of chemical potential 
over which $N$ has a fixed even (odd) value. 

The limits of weak and strong Kondo coupling are readily analysed.\cite{Pereira} 
See Fig. (\ref{fig:steps}). 
When $J=0$, finite width steps only occur for $N$ odd since the energy 
levels in the quantum wire are doubly occupied and a single electron resides 
on the quantum dot. In the opposite limit of strong Kondo coupling, 
finite width steps only occur for $N$ even.  One electron is removed from 
the Fermi sea to screen the spin and the remaining electrons doubly 
occup free fermion energy levels, appropriately $\pi /2$ phase shifted. Thus $R=\infty$ 
at zero coupling and $R=0$ at strong coupling. Again it is possible to 
calculate the corrections to these limits at leading order in 
renormalization group improved perturbation theory at weak coupling 
and Fermi liquid theory at strong coupling. At weak coupling we find:
\be {1\over R}\approx (3/2)[\lambda_0+\lambda_0^2 \ln (L/c)]+\ldots \label{WC}\ee
As for the persistent current, we see that the length of the quantum wire 
cuts off the growth of the Kondo coupling, at $T=0$ and we may write:
\be {1\over R}\to (3/2)\lambda (L)\approx {3\over 2\ln (\xi_K/L)},\ \  (L\ll \xi_K).\ee
In the other limit, first order perturbation theory in the Fermi liquid 
interaction of Eq. (\ref{HFL}) gives:
\be R\to {\pi \xi_K\over 4L},\ \  (L\gg \xi_K).\label{SC}\ee
Again, we see that $R$ appears to be a scaling function of $\xi_K/L$ only. 

\begin{figure}[!ht]
\begin{center}
\includegraphics[height=7cm]{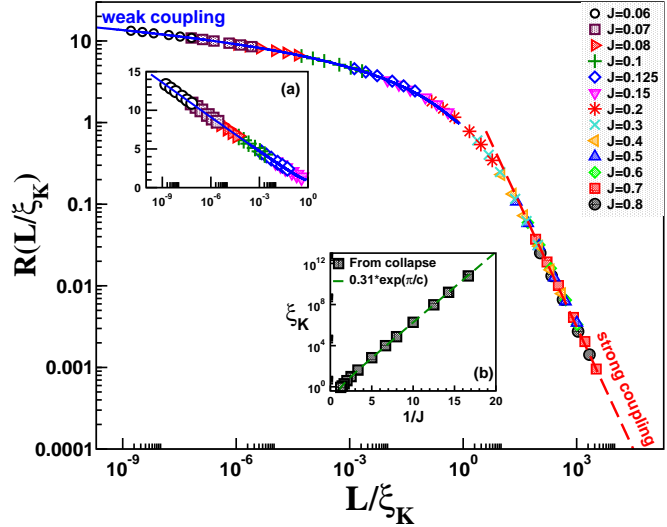}
\end{center}
\caption{Universal ratio of even and odd charge steps, $R=\delta\mu^{\rm{o}}/\delta\mu^{\rm{e}}$,
 as a scaling
function of $\xi_{K}/L$ from [\onlinecite{Pereira}]. Bethe ansatz results\cite{Pereira} 
obtained for various systems sizes ($N_0=51,~101,~201,~501,~1001,~2001$) and 13 different values of the Kondo exchange $J$, indicated by different symbols.
 For each value of $J$, the system lengths $L$ have been rescaled $L\to L/\xi_K(J)$ 
in order to obtain the best collapse of the data using the strong coupling curve 
Eq.~(\ref{SC}) (dashed red line) as a support for the rest of the collapse. 
The weak coupling regime for $L\ll\xi_K$, enlarged in the inset (a), is described
 by the weak coupling expansion Eq.~(\ref{WC}) with ${\rm{constant}}\simeq 0.33$
 (continuous blue curve).
Inset (b): The Kondo length scale, extracted from the universal data collapse of
 the main panel (black squares), is described by theexpected exponential  
dependence on Kondo coupling (dashed green line).}
\label{fig:RBA}
\end{figure}

For this case, we can check this scaling hypothesis, and determined the 
function $R(L/\xi_K)$ to high precision for all $L/\xi_K$ by using 
the Bethe ansatz solution.\cite{Andrei,Weigmann} The result is shown in 
Fig. (\ref{fig:RBA}).  Excellent agreement is obtained with the 
asymptotic weak ($L\ll \xi_K$) and strong ($\xi_K\ll L$) coupling limits.  
Weak particle-hole symmetry breaking has a negligible effect on these results, 
since it simply shifts the entire charge staircase by a constant.\cite{Pereira}

\subsubsection{Including Coulomb interactions in the quantum wire\label{sec:Coulomb}}
So far, we have ignored the effect of bulk Coulomb interactions, only considering 
the Coulomb interaction on the quantum dot, in the Anderson impurity model, which 
leads to the Kondo interaction. This neglect may be justified in many cases, based on Landau 
Fermi liquid theory and the fact that the Kondo interaction just involves 
low energy electron states. However, it is certainly not justified when the Kondo 
screening takes place in a 1D quantum wire.  In this case, Fermi liquid theory breaks down, 
being replaced by Tomonaga-Luttinger Liquid (TLL) theory. In 1D, Coulomb interactions lead to 
major changes in the interaction with an impurity. It turns out however, that these changes 
are minimized when the magnetic impurity is at the end of a quantum wire, so that the 
conclusions of the previous sub-sub-section will not be modified too drastically. 

We modify the model of Eq. (\ref{H0QW}) and (\ref{HKQW}) by adding screened bulk 
Coulomb interactions.  For example, we could consider the Hubbard model with 
only on-site repulsion:
\be H_c=U\sum_j\hat n_{j\uparrow}\hat n_{j\downarrow}\ee
where $\hat n_{j\alpha}$ is the number operator for electrons of spin $\alpha$ at site $j$. 
We expect our results to apply to more realistic models, not based on tight-binding 
approximations and with longer range screened Coulomb interactions. 
We begin by writing the 1D electron annihilation operators in terms of left and right movers, 
similar to what we did for the s-wave projected 3D theory in Eq. (\ref{3D1D}).  
\be c_j\approx e^{ik_Fj}\psi_R(j)+e^{-ik_Fj}\psi_L(j).\ee
Here $\psi_{R/L}$ vary slowly on the lattice scale, containing 
Fourier modes of the $c_j$'s in narrow bands around $\pm k_F$, appropriate for 
the low energy effective theory. 
The Coulomb interactions reduces, in this low energy approximation, to four different 
terms which can be conveniently written in terms of charge and spin densities (or currents) 
for left and right movers:
\be J_{R/L}\equiv \sum_\alpha \psi^{\dagger}_{R/L\alpha}\psi_{R/L\alpha},\ \ 
\vec J_{R/L}\equiv \sum_{\alpha \beta}\psi^{\dagger}_{R/L\alpha}{\vec 
\sigma_{\alpha \beta}\over 2}\psi_{R/L\beta}.\ee
Three of the Coulomb interaction terms are proportional to $(J_L^2+J_r^2)$, 
$J_LJ_R$ and $(\vec J_L^2+\vec J_R^2)$. To make further progress, we bosonize the 
fermions, and introduce separate bosons for the charge and spin degrees of freedom. 
Upon doing this, the first two interaction terms only involve the charge boson. 
They change the velocity, $v_c$ for charge excitations and also rescale 
the boson field by the Luttinger parameter for the charge sector, $K_c$.  The third 
term changes the velocity, $v_s$ for spin excitations. Importantly, all of these 
terms leave a non-interacting model of two massless bosons, for charge and spin. 
The fourth Coulomb interaction (exchange) term is written explicitly as:
\be H_{ex}=-2\pi g_0v_s\int dx \vec J_L\cdot \vec J_R,\ee
where the coupling constant, $g_0$, is proportional to the strength of the Coulomb interactions. 
This can be written entirely in terms of the spin boson operators but corresponds to a 
non-trivial interaction. The exchange interaction, $g_0$ is marginally irrelevant, 
its effects getting weaker at lower energy, so that asymptotically, free boson 
behaviour is obtained. Note that this is the opposite to the behaviour of the Kondo coupling 
which gets stronger at lower energies. We assume the system is not at half-filling so that we 
can ignore Umklapp interactions. 

An important consideration for the charge steps in the wire-dot system we are considering 
is the boundary conditions on the low energy fields, $\psi_{R/L}$ at $j=1$, the end 
of the chain next to the quantum dot. In the limit, $J=0$ these are simply:
\be \psi_L(0)+\psi_R(0)=0.\ee
These can be seen, for example, by adding a ``phantom site'' at $j=0$ and requiring 
that $c_0=0$. Since the low energy fields $\psi_{R/L}$ vary slowly over one lattice spacing, 
the Kondo interaction can be written in terms of $\vec J_R(0)$ only:
\bea c_1&\approx& \psi_R(0)e^{ik_F}+\psi_L(0)e^{-ik_F}\approx 2i\sin k_F \psi_R(0)\\
c_1^\dagger {\vec \sigma\over 2}c_1&\approx& 4\sin^2k_F\vec J_R(0).
\eea
Since $\vec J_R$ can be expressed in terms of 
the spin boson only, the Kondo interaction only involves the spin degrees of freedom. 
This is {\it not} the case if the impurity interacts with the quantum wire far from 
its end.  In that case, we must use the bosonization expression:
\be \psi^\dagger_L(j){\vec \sigma \over 2}\psi_R(j)\approx e^{2ik_Fj}\exp [i\sqrt{2\pi K_c}\phi_c]\vec n\ee
where $\vec n$ can be expressed in terms of the spin boson. Thus, in this situation the 
Kondo interaction mixes the charge and spin sectors leading to quite different physics. 
Here, we have deliberately considered the case of the end-coupled spin in order to 
preserve as much as the physics of the usual Kondo model as possible. 

The charge staircase is determined by the ground state energy, as a function of number of particles. 
In the case where we set to zero both the Kondo interaction, $\lambda_0$ and the bulk marginal coupling, $g_0$, 
this is:
\be E_0-\mu N={\pi v_c\over 4K_cL}(N-N_0)^2 +{\pi v_s\over L}s^2\ee
where $s$ is the spin quantum number of the ground state: 
$0$ for $N$ odd and $1/2$ for $N$ even. (Recall that we include the electron 
on the decoupled quantum dot in our definition of $N$. Here $s$ is the spin of the electrons 
in the wire, not including the quantum dot.) Including both exchange and 
Kondo interactions, it is convenient to write:
\be E_0-\mu N={\pi v_c\over 4K_cL}(N-N_0)^2 +{\pi v_S\over L}f_s(\lambda_0,g_0,L)\ee
where $s=0$ or $1/2$ for $N$ odd or even and $f_s$ are some functions to be determined. 
Then defining:
\be f\equiv f_{1/2}-f_0\label{f01/2}\ee
it is easily seen that the ratio of step widths is:
\be \tilde R ={1+uf\over 1-uf}\label{tildeR}\ee
where 
\be u\equiv v_sK_c/v_c.\ee
Coulomb interactions made $v_c<v_s\approx v_F$ and also reduce $K_c$ from its non-interacting 
value of $1$.  Both these effects reduce $u$ from its 
non-interacting value of $1$ thus bringing $\tilde R$ closer to 1. For cleaved edge 
overgrowth quantum wires $u\approx .5$ and for carbon nanotubes $u\approx .1$. 

In lowest order perturbation theory,
\be f\approx 1-{3\over 2}g(L)-3\lambda (L),\ \  (L\ll \xi_K)\ee
Here we have replaced both the exchange and Kondo couplings by their renormalized values at scale $L$. 
The exchange interaction lowers the ground state energy when the wire (excluding the quantum dot) has an $s=1/2$ 
ground state, corresponding to $N$ even. The effective exchange coupling is:
\be g(L)\approx 1/\ln (L/L_1)\ee
where $L_1$ is a short length scale, of order $k_F^{-1}$ or smaller. The exchange 
interactions actually modify the $\beta$-function for the Kondo coupling:
\be {d\lambda \over d\ln L}=\lambda^2+g\lambda +\ldots \ee
(On the other hand, the Kondo interaction doesn't affect the renormalization of $g(L)$ 
since the Kondo interaction only occurs at the end of the wire and the 
exchange interaction occurs throughout the wire.) Note that the exchange 
interaction causes the Kondo coupling to grow faster as the length scale is 
increased and consequently reduces $\xi_K$ for a given bare Kondo coupling:
\be \xi_K\propto \exp [\sqrt{a/\lambda_0+b}]\ee
for some constants $a$ and $b$ depending on $g_0$. 
For long lengths, $L\gg \xi_K$, we may again use Fermi liquid theory.  Note that the 
effective exchange coupling, $g(L)$ continues to get smaller as we increase $L$ so 
we may continue to treat it perturbatively. When $\xi_K\ll L$ however, one 
electron is removed from the Fermi sea to screen the spin of the quantum dot so that 
the remaining nearly free fermions system has one fewer electron.  This switches 
the $s=0$ and $s=1/2$ ground states and changes the sign of the term in $f$ 
linear in $g(L)$:
\be f \approx -1+{3\over 2} g(L)+{\pi \xi_K\over 2L},\ \  (\xi_K\ll L).\ee
In general we may write:
\be f=f(\xi_K/L,g(L)).\ee
The dependence on $g(L)$ destroys the universal scaling with $L/\xi_K$ but 
in practice this effect can be small since $g(L)$ tends to be small and slowly 
varying. 

\begin{figure}[!ht]
\begin{center}
\includegraphics[height=7cm]{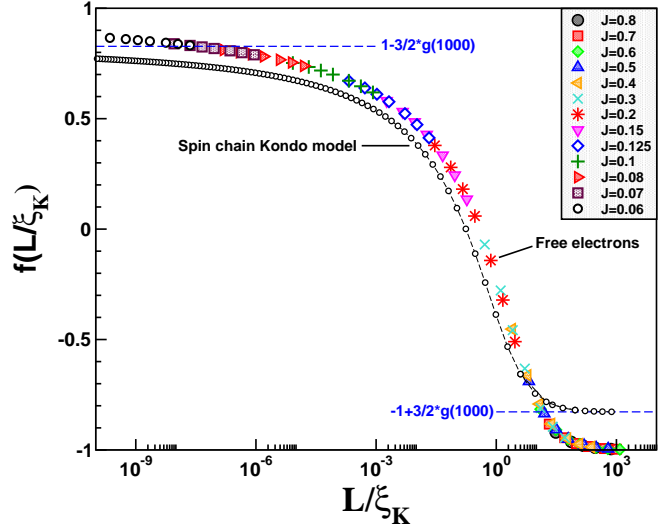}
\end{center}
\caption{Effect of the marginally irrelevant bulk operator on the scaling function
 $f(\xi_K/L,g(L))$ calculated from the Bethe ansatz from [\onlinecite{Pereira}]. The symbols correspond to the
 noninteracting case, $g(L)=0$, as in Fig. \ref{fig:RBA}. The black circles are 
obtained\cite{Pereira}
 with the Bethe ansatz solution of  the spin chain Kondo model for
 $L=1000$ ($g(1000)\simeq 0.115$).}
\label{fig:SCKM}
\end{figure}

Since the Kondo effect for a Luttinger liquid, with the impurity spin end-coupled, 
only involves the spin degrees of freedom, it is essentially unchanged if 
the charge degrees of freedom are gapped, as would occur in the Hubbard model 
at half-filling. In fact, we may take the large $U$ limit at half-filling 
to obtain the S=1/2 Heisenberg antiferromagnetic chain with one weakly coupled 
impurity spin at the end. The only important effect of this large $U$ limit 
is that the bare exchange coupling, $g_0$ becomes O(1). This impurity model has been 
solved via Bethe ansatz methods\cite{Frahm} and this allows an exact calculation\cite{Pereira} 
of the function $f(L/\xi_K,g(L))$ for a particular (large) $g_0$. The results 
are shown, for $L=1000$ and a range of Kondo couplings,
 in Fig. (\ref{fig:SCKM}).  The expected asymptotic limits, $\pm (1-3g(1000)/2)$ are obtained 
for $L/\xi_K\ll 1$ and $\gg 1$. Note that, even for this large bare exchange coupling, 
the deviations from the model with no bulk Coulomb interactions is quite small. 
For conducting quantum wires, we expect $g_0$ to be considerably smaller and 
the deviations to be even less. Thus the main effect of the Coulomb interactions 
is to reduced $\tilde R$ due to the parameter $u$ in Eq. (\ref{tildeR}). 
If $R$ is the step width ratio in the non-interacting model, we may write it 
in the interacting case as approximately:
\be \tilde R(\xi_K/L)={(1-u)+(1+u)R(\xi_K/L)\over (1+u)+(1-u)R(\xi_K/L)}.\ee
This varies from $\tilde R(\xi_K/L)=R(\xi_K/L)$ at $u=1$ (no Coulomb interactions) to 
$\tilde R(\xi_K/L)=1$ at $u=0$ (strong Coulomb interactions). Varying $\xi_K/L$ at 
fixed $u$, $\tilde R$ goes from $(1+u)/(1-u)$ at $L\ll \xi_K$ to 
$(1-u)/(1+u)$ at $\xi_K\gg L$. Coulomb interactions weaken the even-odd alternation 
in the step widths but only trivially modify the dependence on $\xi_K/L$. 

\subsection{Transport through a finite wire containing a quantum dot\label{sec:transport}}
\begin{figure}[!ht]
\begin{center}
\includegraphics[height=3.5cm]{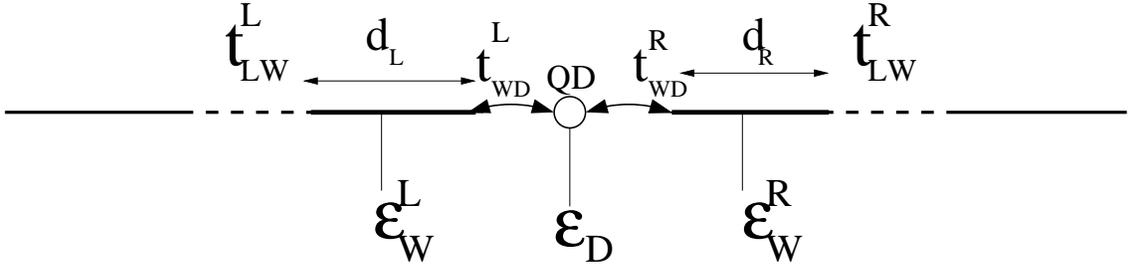}
\end{center}
\caption{Quantum dot embedded in finite length quantum wire. $\epsilon_D$
  and $\epsilon_W^{L/R}$ control respectively the dot and wires gate voltage.  Here we only 
  consider the parity symmetric case whre all ``$L$'' and ``$R$'' quantities are equal.}
\label{fig:QDcond}
\end{figure}

The previous examples have involved closed systems, making the measurement of the Kondo 
effect challenging. Most experiments observing the Kondo effect in quantum dots involve 
transport measurements; the systems are ``open'',  connected to external leads. A natural 
way of observing finite size effects in a transport measurement is to embed 
a quantum dot, in the Kondo regime, in a finite length quantum wire which is 
weakly connected to external leads by tunnel junctions,\cite{Cornaglia,Simon2} as illustrated in 
Fig. (\ref{fig:QDcond}).  Assuming for simplicity, perfect parity symmetry, 
the Hamiltonian is:
\be H=H_L+H_W+H_{LW}+H_{K}\ee
where $L$ labels the leads, $W$ labels the quantum wire and $K$ the Kondo interaction. 
Here:
\bea 
H_{L}&=&-t\left[\sum_{j=-\infty}^{-L-2}+\sum_{j=L+1}^{\infty}\right](c^\dagger_{j}c_{j+1}+h.c.)\\
H_{W}&=&-t\left[\sum_{j=-L}^{-2}+\sum_{j=1}^{L-1}\right](c^\dagger_{j}c_{j+1}+h.c.)
+\epsilon_W\left[\sum_{j=-L}^{-1}+\sum_{j=1}^{L}\right]c^\dagger_jc_j\\
H_{LW}&=&-t_{LW}(c^\dagger_{-L-1}c_{-L}+c^\dagger_Lc_{L+1}+h.c.)\\
H_K&=&J(c^\dagger_{-1}+c_1^\dagger ){\vec \sigma \over 2}(c_{-1}+c_1)\cdot \vec S.\eea
Here spin indices are suppressed and we have ignored a potential scattering term, as 
well as Coulomb interactions in the wire, for simplicity. Note that we have included 
a chemical potential term, $\epsilon_W$ for the wire. When $t_{LW}=0$ the 
wire + dot system is closed and we expect that the length, $L$, of the wire 
will act as an infrared cut-off on the growth of the Kondo coupling, leading to 
scaling in $\xi_K/L$ as discussed in the previous two sub-sections. What happens 
when the wire is connected to the external leads?  When $\xi_K\ll L$, the 
Kondo screening cloud fits inside the leads and, at low $T\ll T_K$, we therefore expect 
perfect transmission through the quantum dot. The conductance is then simply that 
of a finite length non-interacting wire, $W$, connected by tunelling amplitude $t_{WL}$ 
to the external leads. The conductance is then easily calculated by the Landauer formula 
which expresses it in terms of the transmission probability for an electron at the Fermi 
energy to pass through the wire. In this case the Kondo temperature, $T_K$ is unaffected 
by the finite wire length, $L$. On the other hand, when $L\ll \xi_K$ the Kondo 
screening cloud must leak into the external leads. In fact, we might then think of the 
quantum wire plus dot as acting like a large quantum dot.  (But note that, for simplicity,  we haven't 
included an associated charging energy in the Hamiltonian.) 

\begin{figure}[!ht]
\begin{center}
\includegraphics[height=6cm]{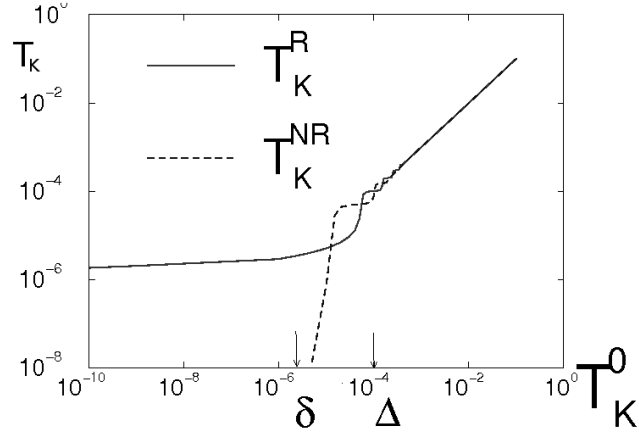}
\end{center}
\caption{The curves\cite{Simon2} represent $T_K=f(T_K^0)$ 
in a log-log scale keeping the same values for $\Delta_n, \delta_n$ for $\epsilon_W$ on
resonance (plain curve which becomes almost flat at low $T_K^0\ll \delta_n$),
and $\epsilon_W$ off resonance (dashed curve which drops sharply at low $T_K^0$).
Both curves coincide at $T_K^0>\Delta_n$.}
\label{fig:TKROR}
\end{figure}

\begin{figure}[!ht]
\begin{center}
\includegraphics[height=6cm, angle=-90]{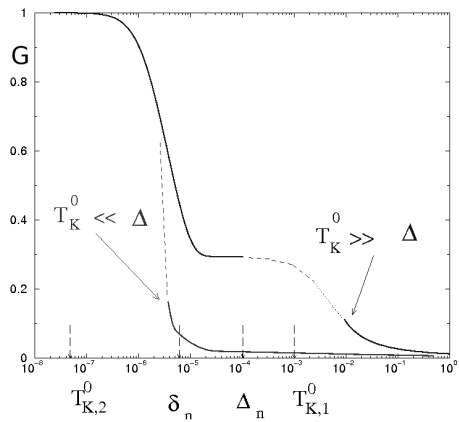}
\end{center}
\caption{On-resonance conductance\cite{Simon2} as a function  of temperature in a symmetric device 
(assuming  $\epsilon_W$ is on resonance) for both cases 
$\Delta_n\ll T_{K,1}^0$ (right blue
curve) and $\Delta_n\gg \delta_n\gg T_{K,2}^0$ (left  red curve), from [\onlinecite{Simon2}]. 
The curves in plain style correspond to the perturbative
calculations plus the Fermi liquid result for the first case only. 
We have 
schematically interpolated these curves (dotted lines) where 
neither the perturbative nor the
Fermi liquid theory applies. }
\label{fig:GWD}
\end{figure}

Ignoring the Kondo interactions, the weak tunnelling between wire and leads, $t_{LW}$, 
leads to a broadening of the 
discrete spectrum of the wire. That is to say, the local density of states, ocurring 
in the Kondo interaction, has peaks at the energies given by the discrete spectrum of 
the wire, with spacing $\Delta = \pi v_F/L$ and broadenned by an amount $\delta \propto t_{LW}^2$. 
We assume that $\delta \ll \Delta$.  
When $L\ll \xi_K$, it was found that the conductance of the device depends very sensitively, 
at low temperatures, on the precise value of $\epsilon_W$, which  shifts 
the broadened energy levels of the wire, relative to the Fermi energy, moving them on or off resonance. 
The renormalization of the Kondo coupling may be calculated straightforwardly in second order,
using this density of states. Off resonance, there is little renormalization of the Kondo coupling 
down to exceedingly low energy scales. On resonance, there is again little renormalization until 
the effective band width is reduced to $\delta$, the width of the resonance peak. At that 
stage the large density of states leads to a rapid growth of the effective Kondo coupling.  
In this situation it is convenient to define an effective Kondo temperature, $T_K$ which 
depends on the length, $L$, of the wire and also on the gate voltage applied to the wire,
 $\epsilon_W$. The non-resonant effective Kondo temperature, $T^{\hbox{NR}}_K\approx T_K$ 
when $T_K\gg \Delta$, i.e. when the screening cloud fits inside the wire. However, 
$T^{\hbox{NR}}$ drops rapidly to zero as $T_K$ is reduced below $\Delta$.  On the other hand, while 
the on-resonance Kondo temperature, $T^{\hbox{R}}_K$, also tracks $T_K$ when $T\gg \Delta$, 
as $T_K$ is reduced further, $T^{\hbox{R}}_K$ levels off at a value of order $\delta$, the 
width of the resonance controlled by $t_{LW}$, as illustrated in Fig. (\ref{fig:TKROR}). 
Correspondly, the on-resonance conductance through the device, when $\delta \ll T_K\ll \Delta$, 
stays very small down to temperatures of order $\delta$ before shooting up towards the unitary limit, 
quite unlike the behaviour in the case $\Delta \ll T_K$, when the conductance approaches the 
unitary limit as the temperature is lowered below $T_K$. This is illustrated in Fig. (\ref{fig:GWD}). 
The dependence of the low $T\ll \Delta$ conductance on the gate voltage, $\epsilon_W$, applied to the wire, 
is also interesting. When the screening cloud doesn't fit inside the wire, $L\ll \xi_K$, there are resonant peaks  
with spacing in $\epsilon_W$ given by the level spacing of the decoupled wires of length $L$, 
$\Delta $.  On the other hand, when the screening cloud fits inside the wire, $\xi_K\ll L$, 
the resonant peaks in low $T$ conductance are separated by $\Delta /2$, the level spacing for 
the doubled wire of length $2L$. 

\subsection{The Kondo Box Model\label{sec:box}}
In 1999, Thimm, Kroha and van Delft introduced\cite{Thimm} a  model of a Kondo 
spin in a finite size system, which they referred to as ``the Kondo box'' model. It behaves 
quite differently than the models reviewed so far in this article, and here we focus on this difference. 
The experimental application that these authors discussed, a single magnetic impurity in a 
disordered ultra-small metallic grain, was rather different than those 
that we have discussed so far. They adopted a simplified Anderson 
impurity model 
in which:
\begin{enumerate}
\item{All energy levels of the unperturbed system (at $J_K=0$) were assumed to be equally spaced in energy
 near the Fermi energy.}
\item{All unperturbed states with energies near the Fermi energy 
were assumed to have equal hybridization with the impurity.}
\end{enumerate}
The corresponding Anderson model is:
\be H=\sum_{i\alpha}\epsilon_ic^\dagger_{i\alpha}c_{i\alpha}+
\epsilon_0\sum_{\alpha}d^\dagger_{\alpha}d_{\alpha}+U\hat n_{d\uparrow}n_{d\downarrow}
+v\sum_{i,\alpha}(c^\dagger_{i\alpha}d_{\alpha}+h.c).\ee
where the $\epsilon_i$ are equally spaced. Taking the large $U$ limit, and ignoring 
potential scattering, the corresponding Kondo model is:
\be H=\sum_{i\alpha}\epsilon_ic^\dagger_{i\alpha}c_{i\alpha}
+J\sum_{i,j}c^\dagger_i {\vec \sigma \over 2}c_j\cdot \vec S\ee
where $\vec S$ is the spin of the magnetic impurity, $\vec S=d^\dagger \vec \sigma d/2$ 
and $J=v^2/U$. The (uniform) energy level spacing, near the Fermi energy was estimated as:
\be \Delta = {\pi^2\over mk_F{\cal V}}\ee
where ${\cal V}$ is the volume of the grain. Note that in the 1D models considered in this 
section, the level spacing is proportional to $1/L$, not $1/{\cal V}$. The behaviour 
of the Kondo box model follows immediately from the discussion of the 1D models earlier in this 
section, if we make the corresponding replacement for $\Delta$. As usual, finite size effects 
become important when $T_k\approx \Delta$, the bulk Kondo temperature is of order the 
level spacing. This implies\cite{Thimm} that grains of order 10 nm. in linear dimension are required 
to see finite size effects, considerably smaller than the estimate of 1 micron in the 1D case. 
The behaviour of this model seems to contradict the naive Kondo screening cloud picture 
since there are essentially no finite size effects even when the linear dimension of the 
grain is much smaller than $v_F/T_K$. 

In fact, the behaviour of this model depends  strongly on the assumptions 1 and 2, above, 
which certainly don't always apply. To make this point it is instructive, as a thought experiment, to consider another 
model of an ultra-small grain which is far less realistic than even the above one. Imagine that 
\begin{enumerate}
\item{The grain is completely clean with no disorder}
\item{The Kondo interaction is short range and spherically symmetric, the standard assumptions in the Kondo model}
\item{The grain is perfectly spherical with the impurity atom exactly at the centre}
\end{enumerate}
In this case, we may expand in spherical harmonics and project onto the s-wave, as usual. The s-wave model 
is essentially the one studied in sub-section \ref{sec:Knight},
 with the impurity spin at one end of a finite length
1D wire. Now finite size effects set in when the length $L$ is of order $v_F/T_K=\xi_K$ as 
discussed above. The essential difference between the two models (or two sets of assumptions) is that 
in the second case most of the states of the metal have zero (or negligible) coupling to the impurity, 
whereas in the first case, all states are assumed to have equal coupling. In the second model 
only the s-wave states couple, a negligible fraction of the total set of states. This happens 
because the higher harmonic wave-functions all vanish at the origin, where the impurity is 
assumed to be located. Note also that the Kondo temperature will be much lower for this spherical model 
since the relevant density of states, appearing in the dimensionless Kondo coupling, scales with 
the volume in the first case but only as the length in the second. 

 The behaviour of the second model is very strongly dependent on the assumptions, 
as can be seen by  relaxing only the third of them. Suppose that we 
allow the surface of the grain to have a non-spherical shape. Then the usual 
wave-functions, involving products of spherical harmonics and spherical Bessel functions, 
are no longer eigenstates of the non-interacting model. If we insist on using this basis, 
then the impurity only interacts with the s-wave as usual.  However, the reflections of the spherical harmonic 
wave-functions off the walls of the grain mix these wave-functions together. Actually, analogous phenomena 
occur in the 1D models considered earlier in this section if we take more realistic models of the 
quantum wires with several channels,\cite{Simon1,Simon2}
 corresponding to transverse wave-functions. Consider, for example, 
the model in Sec. \ref{sec:steps} of a quantum dot at the end of a finite length quantum wire, which has 
several channels. The quantum dot hybridizes with some linear combination of 
channels in the wire. We can always adopt a basis such that one basis state is the 
one hybridizing with the quantum dot. However, these states are mixed by scattering off the 
other end of the wire (away from the quantum dot). A simple, intuitive picture, is that 
the screening cloud wave-function is now a linear superposition of the various channels, 
or the screening cloud folds back upon itself many times inside the wire, due to the 
reflections off the ends. If the wire has $N$ channels, its effective length could be as long as $LN$. 
An important point is that embedding a Kondo spin in a finite system typically destroys 
one of the key simplifying features of the Kondo model - that the impurity spin only couples to 
one channel of unperturbed eigenstates (eg. the s-wave). The only case where we can be sure this 
assumption is still valid is 
when the finite system is an ideal single channel 1D system. 

 Depending on various assumptions about the grain shape, 
magnetic impurity location and disorder, one could imagine the Kondo temperature 
at which finite size effects set in ranging 
anywhere from order $\pi v_F/L$ to order $\pi^2/mk_F{\cal V}$. In all cases, we expect that 
the condition for finite size effects to be important is that $T_K$ be of order the finite 
size gap, suitably weighted by the hybridization of the corresponding wave-functions with the impurity. 
Only in single-channel situations does the simple Kondo screening cloud picture apply. 

Note that this complication makes observation of the screening cloud relatively easier in bulk situations 
where there are no boundaries to mix the hybridizing wave-function with the other channels. 
However, even there, it seems likely that scattering off other Kondo impurities might have a similar effect. 

\section{Conclusions\label{sec:conclude}}
In the standard Kondo model, describing a single magnetic impurity in an infinite host, the length scale 
$\xi_K=v_F/T_K$ appears as a characteristic crossover length scale for many (perhaps essentially 
all) physical observables. Some intuition about the nature of the crossover is provided by 
imagining that, in the ground state, the impurity spin forms a spin singlet with a single conduction electron, 
whose wave-function is constructed out of states within 
energy $T_K$ of the Fermi surface and hence is spread out over a distance of order $\xi_K$. 
One must be careful to only use this picture at low $T\ll T_K$ and even then it sould be 
used with caution.  Naively, we might expect to see weak coupling behaviour, even at $T=0$, at 
distances $r\ll \xi_K$ and strong coupling behaviour, outside the screening cloud, at $r\gg \xi_K$. 
While this seems to apply to the density oscillations and static impurity spin correlations, the 
first of these expectations turns out be be wrong, or at least to need some modifications, in the 
case of the Knight shift. The crossover effects at $\xi_K$ appear to be very hard to observe 
experimentally, precisely because they occur at such long distances. In a D-dimensional sample, 
they are suppressed by a factor of $1/r^D$ (or $1/r^{D-1}$) making their observation more 
favourable in systems of reduced dimensionality. 

In a mesoscopic single-channel 1D systems with a component of finite length, $L$, there is again 
typically a cross-over in physical quantities when $\xi_K$ is of order $L$. In all cases, weak coupling behaviour occurs when 
$L\ll \xi_K$ so that the screening cloud doesn't ``fit'' inside the device and strong coupling behaviour 
occurs when $\xi_K\ll L$ so that finite size effects become unimportant. Except in ideal circumstances, 
mesoscopic systems do not have only a single channel.  While this is unimportant for the Kondo effect in bulk, 
it can have important consequences for the observation of the Kondo screening cloud in a mesoscopic device. 
For an $N$-channel system, the effective length ``seen'' by the Kondo screening cloud may be as large as $LN$.

Throughout most of this paper we have ignored Coulomb interactions between the conduction electrons, 
only taking them into account at the magnetic impurity or quantum dot. While this may be justified at low $T$
by Landau's Fermi liquid theory, for D=2 or 3, it would be necessary for the inelastic scattering length 
to exceed $\xi_K$ for our analyses to apply. In the D=1 case, such interactions are important, even at $T=0$. 
We analysed them only in a particularly simple case, the quantum dot end-coupled to a single channel quantum wire. 
While they somewhat reduced effects associated with the Kondo screening cloud, we concluded that  
the effects are still present and qualitatively unchanged. 

Another effect which can interfere with observation of the screening cloud is non-magnetic disorder. We only 
touched on this briefly in Sec. \ref{sec:box}, is our review of the ``Kondo box'' model. In addition a finite density 
of magnetic impurities further complicates the situation. 

Throughout essentially all of this review, we have considered only the Kondo model, not the more realistic 
Anderson impurity model. We expect similar crossover effects to also occur in that case but with 
another characteristic length scale entering, where the crossover from the free orbital to (unstable) local 
moment fixed point occurs. The Kondo crossover can only be observed if this other crossover occurs 
at a significantly shorter length than $\xi_K$. 

After many years of theoretical and theoretical investigations, the Kondo screening 
cloud, as predicted by the basic Kondo model, remains undetected experimentally. This is likely due to a
combination of experimental difficulties, and limitations of the basic Kondo model. It is to be hoped 
that further experimental and theoretical progress will eventually bring this long search to a happy conclusion.

\section*{Acknowledgements}
I would like to thank all my collaborators in this work.  In chronological order: Erik S\o rensen, 
Victor Barzykin, Pascal Simon, Rodrigo Pereria, Nicolas Laflorencie, Bert Halperin, 
Hubert Saleur and Lazlo Borda. I have also benefited from discussions on this topic with 
Jan von Delft, Junwu Gan, Nikolai Prokof'ev and Fred Zawadowski. This research was supported in part 
by NSERC and CIfAR.

\end{document}